\RequirePackage{pdf14}
\documentclass[letterpaper, 10 pt, conference]{IEEEtran}
\usepackage[utf8]{inputenc} %required for texlive-2016

\usepackage{xcolor}
\usepackage{amsmath}
\usepackage{amssymb}
\usepackage{pgfplotstable}
\usepackage{float}
\usepackage{eurosym}
\usepackage{epsfig}
\usepackage{textcomp}
\usepackage{todonotes,cleveref}
\usepackage{mathtools}
\usepackage{color}
\usepackage{placeins}
\usepackage{tabularx}
\usepackage{dsfont}
\usepackage{graphicx}
\usepackage{lipsum}
\usepackage{footnote}
\usepackage{caption}
\usepackage{subcaption}
\usepackage[noend]{algpseudocode}
\usepackage{algorithm}
\usepackage{algorithmicx}
\algrenewcommand\algorithmicrequire{\textbf{Input:}}
\algrenewcommand\algorithmicensure{\textbf{Output:}}
\usepackage[obeyspaces]{url}

\makeatletter
\def\tagform@#1{\maketag@@@{\normalsize(#1)\@@italiccorr}}
\makeatother

\DeclareMathOperator*{\argmax}{argmax}

\renewcommand{\S}{\mathcal{S}}
\renewcommand{\P}{\mathrm{P}}
\newcommand{\A}{\mathcal{A}}
\newcommand{\R}{\mathcal{R}}
\newcommand{\E}{\mathbb{E}}
\newcommand{\pim}{\pi}
\renewcommand{\H}{\mathcal{H}}

\graphicspath{ {./figs/} }

\title{\LARGE \bf Spatial Positioning Token (SPToken) for Smart Mobility}

\makeatletter
\def\footnoterule{\kern-3\p@\hrule \@width 1.5in \kern 2.6\p@}
\makeatother

\author{
    Roman Overko\IEEEauthorrefmark{1}, Rodrigo Ord\'{o}\~{n}ez-Hurtado\IEEEauthorrefmark{2},
    Sergiy Zhuk\IEEEauthorrefmark{2}, \\ 
    Pietro Ferraro\IEEEauthorrefmark{3}, Andrew Cullen\IEEEauthorrefmark{3}, and
    Robert Shorten\IEEEauthorrefmark{3}
    \thanks{
        \IEEEauthorrefmark{1} University College Dublin, Dublin, Ireland. Email: 
        {\tt roman.overko@ucdconnect.ie}
    }
    \thanks{
        \IEEEauthorrefmark{2} IBM Research, Building 3, Damastown Industrial Park, Dublin 15, 
        Ireland. Email: {\tt rodrigo.ordonez.hurtado@ibm.com, sergiy.zhuk@ie.ibm.com}
    }
    \thanks{
        \IEEEauthorrefmark{3} Imperial College London, South Kensington, London SW7 2AZ, 
        United Kingdom. Email: {\tt \{p.ferraro, a.cullen19, r.shorten\}@imperial.ac.uk}
    }
    \thanks{
        \textcopyright 2020 IEEE. Personal use of this material is permitted. Permission from IEEE 
        must be obtained for all other uses, in any current or future media, including 
        reprinting/republishing this material for advertising or promotionalpurposes, creating new 
        collective works, for resale or redistribution to serversor lists, or reuse of any 
        copyrighted component of this work in other works.
    }
}
\IEEEoverridecommandlockouts % to enable \thanks command in ieeeconf document

\begin{document}
\maketitle

\begin{abstract}
    We introduce a permissioned distributed ledger technology (DLT) design for crowdsourced smart 
    mobility applications. This architecture is based on a directed acyclic graph architecture 
    (similar to the IOTA tangle) and uses both Proof-of-Work and Proof-of-Position mechanisms to 
    provide protection against spam attacks and malevolent actors. In addition to enabling individuals 
    to retain ownership of their data and to monetize it, the architecture also is suitable for 
    distributed privacy-preserving machine learning algorithms, is lightweight, and can be implemented 
    in simple internet-of-things (IoT) devices. To demonstrate its efficacy, we apply this framework 
    to reinforcement learning settings where a third party is interested in acquiring information from 
    agents. In particular, one may be interested in sampling an unknown vehicular traffic flow in a 
    city, using a DLT-type architecture and without perturbing the density, with the idea of realizing 
    a set of virtual tokens as surrogates of real vehicles to explore geographical areas of interest. 
    These tokens, whose authenticated position determines write access to the ledger, are thus used to 
    emulate the probing actions of commanded (real) vehicles on a given planned route by ``jumping'' 
    from a passing-by vehicle to another to complete the planned trajectory. Consequently, the 
    environment stays unaffected (i.e., the autonomy of participating vehicles is not influenced by 
    the algorithm), regardless of the number of emitted tokens. The design of such a DLT architecture 
    is presented, and numerical results from large-scale simulations are provided to validate the 
    proposed approach.
\end{abstract}

\section{Introduction}\label{sec:introction}
    Companies such as Facebook, Google, Amazon, Waze and Garmin are just some examples of corporations 
    that have built successful service delivery platforms using personal data to develop recommender 
    systems. While products gleaned from data mining of personal information have without doubt 
    delivered a great societal value, they have also given rise to a number of ethical questions that 
    are causing a fundamental revision on how data is collected and managed \cite{zhang2018big}. 
    Some of the most pressing ethical issues include:
    \begin{itemize}
        \item[1.] preservation of individuals' privacy (including GDPR compliance);
        \item[2.] the ability for individuals to retain ownership of their own data;
        \item[3.] the ability for consumers and regulatory agencies alike to confirm the origin, 
            veracity, and legal ownership of data;  
        \item[4.] protection against misuse of data by malevolent actors.\\
    \end{itemize}

    In this context, Distributed Ledger Technology (DLT) has much to offer. For example, it has been 
    shown in a number of specialized papers (e.g., \cite{zyskind2015decentralizing, liang2017provchain}) 
    that the use of technologies such as blockchain has proven beneficial to alleviate, or even 
    eliminate, some of these above data ethics considerations. Consequently, our objective in this 
    paper is to design one such system. We are particularly interested in developing a DLT-based 
    architecture that supports the design and realization of crowdsourced collaborative recommender 
    systems to sustain a range of mobility applications for smart cities. This objective is challenging 
    for a number of reasons. First, from the perspective of the basic distributed ledger design, 
    the system must support high-frequency microtransactions to facilitate the rapid exchange of 
    information required by the multitude of IoT-enabled devices found in cities. Second, as the DLT 
    must support multiple control actions and recommendations in real time, transaction times should 
    be short. Finally, the ledger architecture should penalize malevolent actors who 
    attempt to spam the system or lie to attack the design of any recommender system based on the DLT.
    \newline
    
    It is also worth noting that wrapping a DLT layer around personal information will fundamentally 
    change the business model of many companies \cite{nowinski2017can}. Big-data corporations currently 
    monetize recorded personal data with no explicit reward returned to the owner of such data (other 
    than personalized recommendations or free access to products in return for the collected data). 
    In a scenario in which such data is no longer available free-of-charge to these corporations, they 
    would need to find alternative solutions to maintain their quality of service. As an example, 
    imagine a company such as Waze that monitors traffic patterns in cities, and which can build a 
    complete picture of the city environment in almost real time via a telematics platform 
    \cite{graaf2018waze}. If in the future all data became privately held and not available in a 
    public manner, the view of the city environment to this company would be highly uncertain. 
    This \emph{disruption} will only be alleviated by purchasing data from many car owners. This 
    scenario raises a fundamental question: {\it how to sample privately held data given some 
    desired level of accuracy (e.g., a minimum quality of service)?} Given this background, an 
    essential requirement is to develop a set of tools that enable large-scale data sets to be sampled, 
    with a specific objective in mind. For instance, in the Waze example, we wish to reconstruct 
    traffic densities, but obtaining data to build specific predictive tools may require a much more 
    refined data collection procedure. We also wish for the sampling to be done in a manner that 
    allows the veracity of the data to be verified, which enables data owners to be compensated in 
    proportion to the quality of the data they provide.
    \newline

    An additional challenge arises from the design of recommender systems itself. In many important 
    applications, the development of complex decision making tools is inhibited by difficulties in 
    interpreting large-scale, aggregated data sets. This difficulty stems from the fact that data sets 
    often represent {\em closed-loop} situations, where actions taken under the influence of decision 
    support tools (i.e., recommenders), or even due to probing of the environment as a part of the 
    model building, do affect the surroundings and consequently the model structure itself. 
    Recently, a number of papers have appeared highlighting the problem of recommender design in 
    closed loop \cite{Lazer2014, Sinha_NIPS17, Shorten_IEEETech16, Bottou, jonathan}. Even in cases 
    when there is a separation between the effect of a recommender and its environment, the problem 
    of recommender design is complex in many real-world settings due to the challenge of sampling and 
    obtaining real-time data at low cost.
    \newline

    In this paper, we bring both of the above problems together in one framework. In particular, 
    we consider the problem of sampling an unknown density representing traffic flow in a city, 
    comprised of secured data points, using a DLT architecture, without perturbing the density 
    through probing actions. Specifically, we will use reinforcement learning (RL) \cite{RN229} 
    to explore the density in order to build a model of the environment. Classical RL is usually 
    not applicable for this purpose in many smart city applications due to long training time, and 
    due to the disruptive effects of probing. However, we shall demonstrate that the use of a DLT-based 
    structure allows us to probe the surroundings without affecting it and also enables individuals 
    to not only retain ownership of their own data, but to potentially be rewarded for contributing 
    to the RL framework.
    \newline

    Our paper is structured as follows. Related work is presented in Section 2. In Section 3, we 
    present a distributed ledger design that can be used in smart mobility applications. We then, 
    in Section 4, illustrate how this ledger can be used to facilitate collaborative reinforcement 
    learning algorithms based on crowdsourced information without giving rise to undesirable 
    closed-loop effects (artificial traffic jams). Finally, simulation results are presented in 
    Section 5, and concluding remarks presented in Section 6.
    \newline

    {\bf Note to reader:} A short version of this paper was an award winner when presented at the 
    International Conference on Connected Vehicles and Expo (ICCVE) 2019 \cite{roman_iccve_2019}.
    The present manuscript extends this basic version significantly in several ways.
    \begin{itemize}
        \item[(i)] All sections of the present paper are more detailed than \cite{roman_iccve_2019}.
            The ICCVE paper is a 6 page manuscript whereas the present manuscript is 14 pages long.
        \item[(ii)] The adaptive Proof-of-Work mechanism described in Section \ref{sec: PoP} of the 
            present manuscript is not used in \cite{roman_iccve_2019}.
        \item[(iii)] In \cite{roman_iccve_2019}, each road link represents a state of a Markov chain. 
            In this work, we merge appropriate road segments into a single state when possible (e.g., 
            there is only one action from a given source link). The advantage of this is a significant 
            reduction in the number of states.
        \item[(iv)] The new road network used in the experimental section is much larger (3663 merged 
            states as opposed to 473 states) than in \cite{roman_iccve_2019}.
        \item[(v)] In this work, we also design the state-action space 
            with the following modifications: (i) each state contains an independent subset of actions, 
            and (ii) the set of actions includes two more actions.
        \item[(vi)] More extensive and realistic simulation results are presented, and all the 
            experiments are new and much more extensive than those in \cite{roman_iccve_2019}.
    \end{itemize}

\section{Related work}
    While our work brings together ideas from many areas, the principal idea presented is the use 
    of a DLT to enable distributed reinforcement learning algorithms that can be applied in 
    crowdsourced environments (while avoiding undesirable closed-loop effects). That being said, 
    the paper is not about crowdsourcing per se, and while we use a routing-based RL example to 
    illustrate our architecture, the paper is also not advocating a particular routing policy.
    \newline

    DLT is a term that describes {\it blockchain} and a suite of related technologies. From a 
    high-level perspective, a DLT is nothing more than a ledger held in multiple places, and a 
    mechanism for agreeing on the contents of the ledger---namely the {\it consensus mechanism}. 
    Since the debut of blockchain in 2008 \cite{nakamoto2008bitcoin}, the technology has been used 
    primarily as an immutable record keeping tool that enables financial transactions based on 
    peer-to-peer trust \cite{puthal2018everything, conoscenti2016blockchain, zheng2017overview, 
    banerjee2018blockchain, yli2016current}. Architectures such as blockchain operate a competitive 
    consensus mechanism enabled via {\it mining} (i.e., Proof-of-Work), whereas architectures such as 
    the IOTA Tangle \cite{wang2018survey} based on other graph structures can facilitate cooperative 
    consensus techniques. In our proposed approach, we use a DLT based on the IOTA Tangle. 
    Our interest in IOTA stems from the fact that its architecture is designed to facilitate 
    high-frequency microtrading. In particular, the architecture places a low computational and 
    energy burden on devices using IOTA, it is highly scalable, there are no transaction fees, and 
    transactions are pseudo-anonymous \cite{popov2017equilibria}. In terms of mobility applications, 
    we note that several DLT architectures have already been proposed. Recent examples include 
    \cite{carnet, towards, bc} and the references therein. To the best of our knowledge, our work 
    is the first using a Directed Acyclic Graph (DAG) structure, namely the IOTA Tangle, to support 
    distributed machine learning (ML) algorithms.
    \newline

    In terms of ML, we borrow heavily from RL, Markov Decision Processes (MDPs), and, in particular, 
    crowdsourced ML. The literature on MDPs and RL algorithms is vast and we simply point the reader 
    to some relevant publications in which some of this work is discussed (see \cite{jonathan, Krumm2008, 
    SimmonsBrowningZhangEtAl2006}). With specific regard to RL and mobility, some applications are 
    presented in \cite{work15, work18, work17, work19, work16}. As in our previous work 
    \cite{roman_cdc_2019}, we exploit the idea of using crowdsourced behavioural experience to 
    augment the training of ML algorithms (see \cite{work16} for an example of multi-agent RL and 
    a recent survey for an overview of this area in \cite{Vaughan2018}). 
    We also note that our approach bears similarities to the concept of {\em virtual trip lines} 
    introduced in \cite{virt_trip_lines} and further developed in \cite{virt_trip_lines_2012}. 
    However, as we shall see in the sequel, our proposed design goes beyond traffic monitoring: 
    it involves a token-passing mechanism, it uses a DLT-based 
    architecture, it does not depend on GPS traces and it uses Blockchain-like ideas to enforce data 
    sovereignty and spamming prevention. Note that some of these objectives are 
    somewhat related to the literature on classical crowdsourcing \cite{rev1_2,rev1_1}, where one of 
    the main goals is to quantify data quality in a crowdsourcing setting. In our situation, the use of 
    DLT is motivated by the advantages that the architecture brings to this domain (crowdsourcing): 
    DLT's built-in consensus mechanism allows, in fact, to both increase quality of data written to 
    ledger, and to prevent spamming. Furthermore, its distributed nature allows the provenance of data 
    to be recorded so that value can later be returned to data owners as the data is used. Also, even 
    though our DLT design uses techniques different to those described in \cite{rev1_2,rev1_3}, those 
    later techniques are in fact complementary to those proposed here. Namely, those techniques can 
    also apply in our setting to data that is already written in the ledger. This can be viewed as a 
    supplementary layer to the existing DLT architecture, since mechanisms that make difficult 
    writing malicious data to the ledger are in fact intrinsic to the DLT approach (as we shall 
    discuss in later sections). We note also that, as mentioned in \cite{roman_cdc_2019}, our work 
    also has strong links to adaptive control \cite{n1}. The idea of augmenting offline models with 
    adaptation is discussed extensively in the recent {\em multiple-models, switching, and tuning} 
    paradigm \cite{n2}.
    \newline

    It is worth mentioning that we are ultimately interested in the design of recommender systems 
    that account for feedback effects in smart city applications. In \cite{florian, bei, arieh}, 
    different information is sent to different agents in an attempt to mitigate closed-loop effects. 
    An alternative, more formal, approach is presented in \cite{jonathan}. There, the authors 
    attempt the identification of a smart city system from closed-loop data sets. Similar issues 
    have drawn interest from various domains including economics~\cite{HalVarian_NASUS17}, recommender 
    systems~\cite{Cosley2003, Sinha_NIPS17}, physiology~\cite{Gollee11}, and control engineering in 
    the context of Smart Cities \cite{Shorten_IEEETech16}.
    \newline

    Finally, while we have strongly emphasized that this paper is not concerned with routing per se, 
    we nevertheless present an application in which RL is used to learn optimal routes in a complex 
    environment. Although routing is not our focus, we do note that this is a very rich area of 
    research. For a flavour of existing work see, for example, Chapter 5 of \cite{booke} for an 
    overview of the literature with specific regard to special vehicles (connected and electric) and 
    the references therein, and the papers \cite{rev2_1,rev2_2,rev2_3}. In principle, the 
    crowdsourced DLT-based architecture can be applied in these situations too.

\section{A distributed ledger for crowdsourced smart mobility - SPToken}\label{sec:dlt}
    \subsection{Design objectives}
\begin{figure}
        \centering
        \begin{subfigure}[b]{1\columnwidth}
            \includegraphics[width=1\columnwidth]{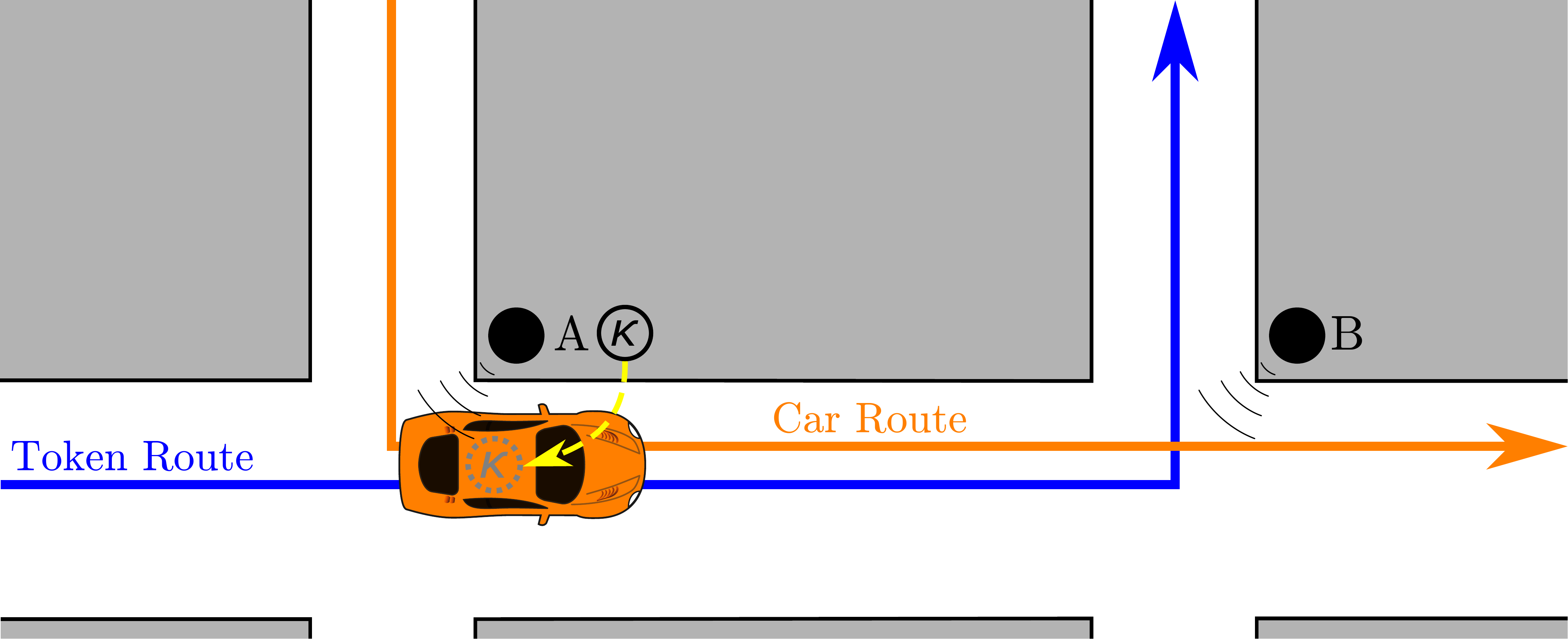}
            \caption{A vehicle passes through a junction $A$ where another car has recently issued some 
                data (this is displayed by a token). This makes the agent eligible to write transactions 
                to the ledger.}
        \end{subfigure}
        \medskip
        \begin{subfigure}[b]{1\columnwidth}
            \includegraphics[width=1\columnwidth]{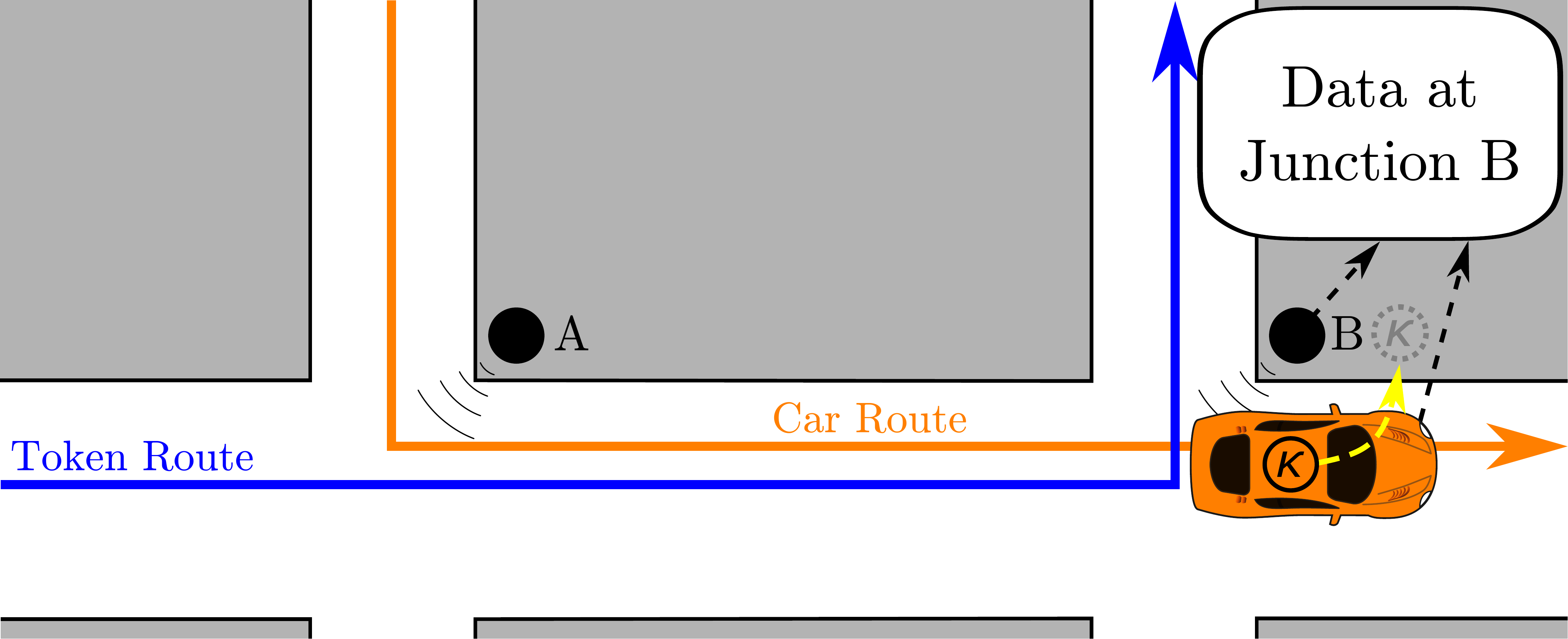}
            \caption{The same vehicle passes through junction $B$. It then writes some data, relative 
                to the road link $\overrightarrow{AB}$, to the ledger and deposits the token so that 
                another vehicle will be able to collect it.}
        \end{subfigure}
        \caption{The sequence to issue new data from vehicles. Here $\kappa$ denotes a token. 
            [Clipart from www.clker.com]}
        \label{Fig: Token}
    \end{figure}

    Our aim is to design a DLT-based system for crowdsourcing in a smart mobility environment. 
    The DLT is nothing more than a ledger used to record data points and their ownership.
    Competing devices, namely cars, are able to write data to the ledger in a manner that ensures 
    reliable and high-quality data are written to the ledger. Note that the ledger records data points 
    and their sovereignty, among other information, all of which can be encrypted through classical 
    methods, so that the owner maintains control over who is allowed to access it. A third party then, 
    possibly via monetary payments to the respective data owners, accesses the ledger and
    uses these data points to run a collaborative algorithm. Value is then returned 
    to individual data owners using some techniques (for example, using the ideas based on 
    Shapley Value \cite{rade}). More specifically, in the present paper we explore how to apply this 
    framework to an RL setting where a third party is interested in 
    acquiring information from vehicles in order to solve an optimization problem. In doing this, we 
    focus on how the DLT is used to enable crowdsourced information to realize the RL algorithm 
    (rather than on monetary reward for each vehicle).
    The underlying idea is to use a set of virtual \emph{tokens} as a proxy to indicate specific 
    geographical points of interest, whose absolute/relative states and surrounding conditions (e.g., 
    position, speed, nearby air pollution) are of significance for dedicated algorithms. 
    In RL algorithms, for example, 
    we are interested in maximizing the expected reward (relative to an objective function) for taking a 
    specific route across a city. To make this process clearer, consider the following example. Figure 
    \ref{Fig: Token} shows an instance of a typical scenario where two road 
    junctions $A$ and $B$ are connected to 
    one another through the road segment $\overrightarrow{AB}$. At time $T_A$, a vehicle updates the 
    ledger with some collected information $x^{T_A}$ (e.g., air pollution level, travel time) by 
    registering at a given visited intersection ($A$, in this example). Intuitively, this can be 
    depicted as if the vehicle leaves a token $\kappa$ with associated information $x^{T_A}$ at 
    junction $A$. Then, a new vehicle passing via junction $A$ and directed to junction $B$ 
    can ``collect'' token $\kappa$ and, as it passes by junction $B$ at time $T_B>T_A$, it updates the 
    ledger with new information $x^{T_B}$ regarding route link $\overrightarrow{AB}$ and the new 
    position of the token. It is noteworthy that in Figure \ref{Fig: Token} a vehicle leaves the 
    token when it deviates from the token route. Thus, a new car that passes by junction $B$ whose 
    immediate future trajectory coincides with the token's route will be able to collect the token 
    and the procedure is repeated for a new road segment.
    \newline

    The concept of using tokens to be deposited at specific locations where measurements are needed 
    perfectly conforms with a DLT-based system. In fact, it is natural to use distributed ledger 
    transactions to update the position of available tokens and register the associated data to the 
    points of interest by using transactions (which can be done, for example, using smart sensors at 
    various junctions linked to digital wallets, as shown in Figure \ref{Fig: Token}). Of course, the 
    design of such a network poses a number of challenges that need to be addressed:
    \begin{itemize}
        \item \emph{Privacy:} In the DLT, transactions are 
            pseudo-anonymous\footnote{\url{https://laurencetennant.com/papers/anonymity-iota.pdf}}.
            This is due to the cryptographic nature of the addressing, which is less revealing than 
            other forms of digital payments that are uniquely associated with an individual 
            \cite{2019arXiv190107302F}. Thus, from a privacy perspective, the use of DLT is desirable 
            in a smart mobility scenario.
        \item \emph{Ownership:} Transactions in the DLT can be encrypted by the issuer, thus allowing 
            every agent to maintain ownership of their own data. In the aforementioned setting, the 
            only information required to remain public is the current ownership of the tokens.
        \item \emph{Microtransactions:} Due to the amount of vehicles in the city environment, and 
            also due to the need for linking the information to real-time conditions (such as traffic ]
            or pollution levels), there is the demand for a fast and large data throughput.
        \item \emph{Resilience to Misuse:} The system must be resilient to attacks and misuse from 
            malevolent actors. Typical examples include double spending attacks, spamming the system, or 
            writing false information to the ledger. All these instances can be greatly limited by a 
            combined use of a consensus system based on Proof-of-Work (PoW) and Proof-of-Position (PoP), 
            which will be described in the next section. 
    \end{itemize}
    
    To meet all the design objectives described above, in the next section we propose \emph{Spatial 
    Positioning Token} (SPToken), a permissioned distributed ledger for smart mobility applications, 
    based on the IOTA Tangle.

    \subsection{The Tangle, Proof-of-Position, and Adaptive Proof of Work}\label{sec: PoP}
\begin{figure}
        \centering
        \includegraphics[width=0.7\columnwidth]{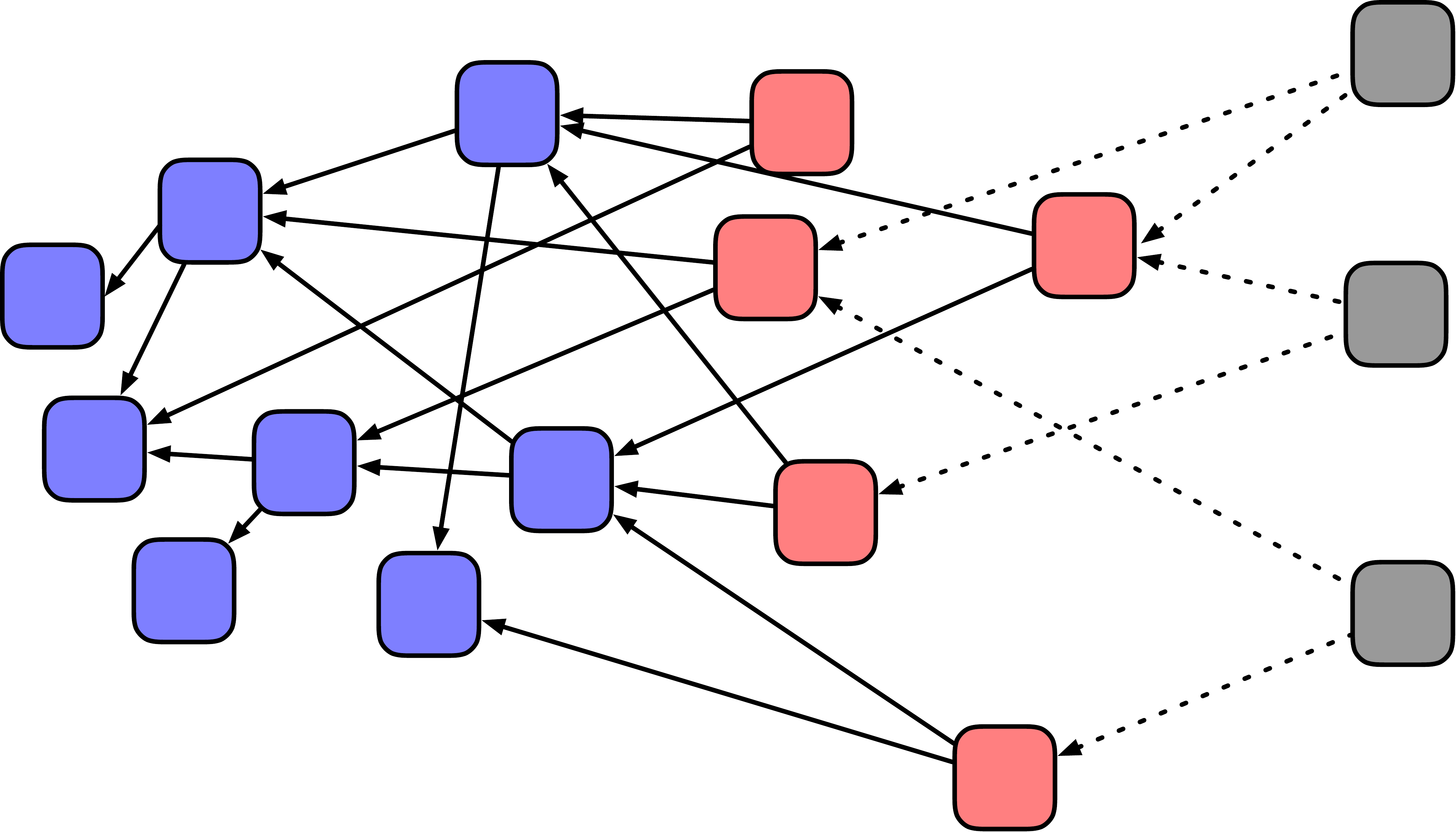}
        \includegraphics[width=0.7\columnwidth]{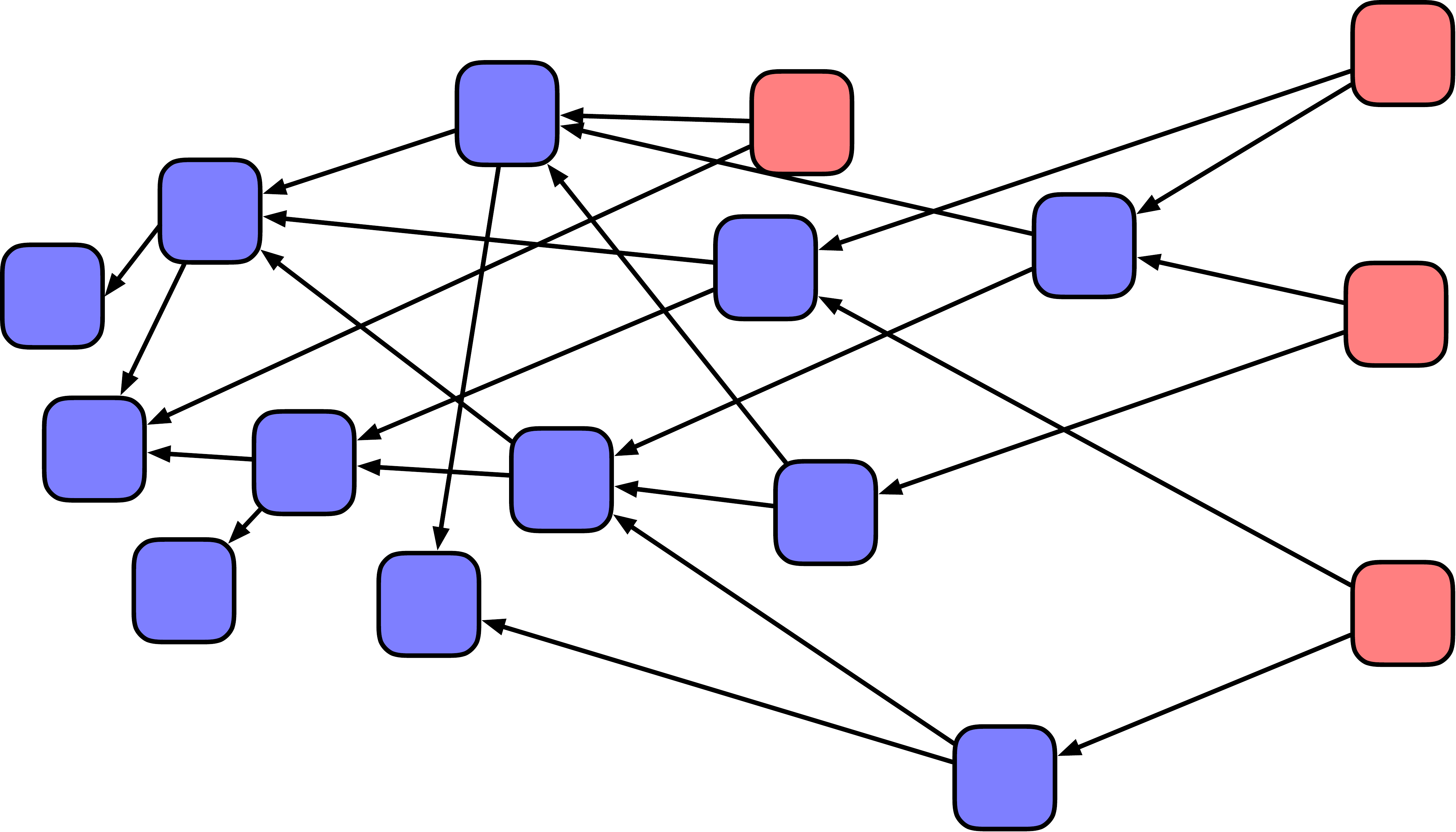}
        \caption{Sequence to issue a new transaction. The blue sites represent the approved 
            transactions and the red ones describe transactions that have not been approved yet. 
            The black edges represent approvals, whereas the dashed ones represent transactions that 
            are performing the PoW in order to approve two unapproved sites.}
        \label{Fig: Tangle}
    \end{figure}

    As discussed above, we are interested in building a particular Tangle-based, DLT architecture that 
    makes use of DAGs to achieve consensus on the shared ledger. A DAG is a finite connected directed 
    graph with no directed cycles. In other words, in a DAG there is no directed path that connects 
    a vertex with itself. The IOTA Tangle is a particular instance of a DAG-based DLT 
    \cite{popov2017equilibria}, where each vertex or \emph{site} represents a transaction, and where the 
    graph with its topology represents the ledger. Whenever a new vertex is added to the Tangle, this 
    must approve a number of previous transactions (normally two). An \emph{approval} is represented by a 
    new edge added to the graph. Furthermore, in order to prevent malicious users from spamming the 
    network, the approval step requires a small PoW. This step is less computationally intense than 
    its blockchain counterpart \cite{banerjee2018blockchain} and can be easily carried out by common 
    IoT devices, but still introduces some delay for new transactions before they are added to the 
    Tangle. Refer to Figure \ref{Fig: Tangle} for a better understanding of this process.
    \newline

    The Tangle architecture has the advantage over blockchain to allow microtransactions without any fees 
    (as miners are not needed to reach consensus over the network \cite{2019arXiv190107302F}), 
    which makes it ideal in an IoT setting as it is described in the previous section. Moreover, the 
    Tangle fits perfectly with the concept of multiple tokens being transferred from one location to 
    another as its DAG structure makes it natural to describe such a process.
    \newline

    Unlike the Tangle, in which each user has complete freedom on how to update the ledger with 
    transactions, the SPToken ledger is designed to have an additional regulatory policy in order to 
    prevent agents from adding transactions that do not possess any relevant data (since transactions 
    are encrypted). Therefore, as a further security measure, SPToken makes use of PoP to authenticate 
    transactions: for a transaction to be authenticated, it has to carry proof that the agent was 
    indeed in an area where a token was present, which is achieved via special nodes called 
    \emph{observers} linked to physical sensors in a city\footnote{A sensor can be a fixed piece of 
    infrastructure, or a vehicle whose position is verified.}. Whenever a participating car passes 
    by an observer that is in possession of a token, a short range wireless connection is established 
    (e.g., via Bluetooth) and the token is transferred to the vehicle's account if the 
    requirements are met (e.g., the immediate vehicle's and token's future trajectories intersect). 
    To deposit the token and to issue a transaction containing data, the agent needs to pass by another 
    observer and establish a short range connection. See Figure \ref{Fig: Token} for a better 
    understanding of this process. This process ensures that vehicles have to be physically at the 
    observation points to be able to issue transactions. Such an additional authentication step makes 
    SPToken a permissioned Tangle (similar to permissioned blockchains \cite{puthal2018everything}) 
    i.e. a DAG-based distributed ledger where a certain number of trusted nodes (the observers) are 
    responsible for maintaining the consistency of the ledger (as opposed to a public one, where 
    security is handled by a cooperative consensus mechanism \cite{popov2017equilibria}). To keep 
    users from hoarding tokens, the latter should be automatically returned if not used after a 
    certain period of time. 
    \newline

    Furthermore, an additional {\it adaptive Proof-of-Work} (aPoW) step can be introduced. The aim of 
    this aPoW step is to strategically vary the target difficulty of the PoW to hinder the writing 
    process for ``bad'' agents that plan to register false information. This is achieved by allowing 
    observers to issue a number of tokens for each desired data point rather than only one. 
    Once each of the vehicles in possession of a token completes a physical PoP step (namely, 
    finishing a road segment), they must also complete a PoW step, the difficulty of which depends on 
    the value of the data they are attempting to write to the ledger. Specifically, vehicles trying to 
    write false data should have to complete a more difficult PoW. The introduction of the aPoW step 
    relies on the assumption that the majority of users will behave honestly and will offer data that 
    is genuine. Moreover, for mobility applications, and thanks to the PoP step, it is reasonable to 
    expect that honest agents taking measurements in the same area, at the same time, will provide 
    similar data (e.g., traffic density, pollution levels).
    \newline

    Consider the event where $N$ users, with writing privileges, each tries to add a data point $x_i$ 
    to the DAG, where $x_i \in \mathbb{R}^{n}, i \in \mathcal{N}$, $\mathcal{N}=\{1,2,\ldots, N\}$. 
    We define $d_w(\cdot, \cdot, \cdot)$ to be the amount of work needed to add a new item to the DAG. 
    This work is the difficulty of solving the aPoW step, which is equal to some fixed value $d_0$ 
    (representing the minimum amount of work required), plus a term that takes into account the average 
    ``distance'' between the data $x_i$ and all the other items being added at that time. This can 
    be expressed as
    \begin{equation}
        d_w(x_i, \overline{\text{x}}, d_0) = d_0 + \alpha \, \left\Vert x_i 
            - \overline{\text{x}}\right\Vert,
        \label{eq:aPoW}
    \end{equation}
    \begin{equation}
        \overline{\text{x}}=\dfrac{1}{N}\sum_{j\in \mathcal{N}}x_j, \nonumber
    \end{equation}
    where $\left\Vert\cdot\right\Vert$ is a norm in $\mathbb{R}^{n}$ and $\alpha >0$ is a constant 
    scaling factor. This mechanism ensures that anyone trying to update the ledger with data that are 
    much different from its competitors will have to spend more computing power than other users in 
    order to complete the aPoW. While the aPoW would be straightforward to implement on a regular DLT, 
    its use requires $x_i$ and $\overline{\text{x}}$ to be known to all users (in order to check if a 
    nonce that satisfies $d_w(x_i, \overline{\text{x}})$ was found). A possible way to circumvent 
    this problem and maintain the ability to encrypt the data is to operate the following 
    privacy-preserving scheme (for each competitor $i$ in parallel):
    \begin{enumerate}
        \item decompose $x_i$ non-uniformly\footnote{If all $x_i$ were split uniformly, the trivial 
            solution $c_1 = c_2 = \dots = c_N$ would result in a violation of the scheme's privacy.} 
            into $N$ \emph{fragments} $x_{ij}$, such that 
            \begin{equation}
                \sum_{j\in\mathcal{N}} x_{ij} = x_i,
                \label{eq:fragmentation}
            \end{equation}
            keep $x_{ii}$ and broadcast all the remaining $N-1$ fragments $x_{ij}$, $\forall i \neq j$, 
            to each other competitor $j\in\mathcal{N}-\left\{i\right\}$;
        \item receive the corresponding $N-1$ fragments, one from each other competitor, calculate 
            $c_i = \sum_{k\in\mathcal{N}}x_{ki}$, and broadcast it to the other 
            competitors\footnote{Note that $\sum_{i \in\mathcal{N}} c_i = \sum_{i \in\mathcal{N}} x_i$.};
        \item based on the received sum of fragments, compute 
            $\overline{\text{x}}=\dfrac{1}{N}\sum_{k\in\mathcal{N}} c_k$;
        \item complete $N$ aPoW steps, with difficulty levels $d_w\left(x_{ik}, 
            \overline{\text{x}}/N, d_0/N\right)$, $k \in\mathcal{N}$. 
    \end{enumerate}
    In step 4), for each one of those aPoW steps, competitor $k$ certifies that competitor $i$ has 
    completed its aPoW and that $x_{ik}$ has been issued individually (to avoid cheating). Note that 
    by placing (\ref{eq:fragmentation}) into (\ref{eq:aPoW}), and since $\left(\cdot\right)$ can be 
    expressed as $\frac{1}{N}\sum_{k\in\mathcal{N}}\left(\cdot\right)$, then we obtain
    \begin{eqnarray}
        d_w(x_i, \overline{\text{x}}, d_0) 
        &=& d_0 + \alpha\left\Vert\sum_{k\in\mathcal{N}}x_{ik} - \overline{\text{x}}\right\Vert
        \nonumber \\
        &=& \sum_{k\in\mathcal{N}}\frac{d_0}{N} + \alpha\left\Vert\sum_{k\in\mathcal{N}}\left(x_{ik} - 
        \frac{\overline{\text{x}}}{N}\right)\right\Vert
        \nonumber \\
        & \leq & \sum_{k\in\mathcal{N}}\left(\frac{d_0}{N} + \alpha\left\Vert x_{ik} - 
        \dfrac{\overline{\text{x}}}{N}\right\Vert\right).
        \label{eq:dw_new}
    \end{eqnarray}
    Using the same notation as in (\ref{eq:aPoW}), we arrive to
    \begin{equation}
        d_w(x_i, \overline{\text{x}}, d_0) \leq 
        \sum_{k \in\mathcal{N}} d_w\left(x_{ik}, \frac{\overline{\text{x}}}{N}, \frac{d_0}{N}\right),
        \label{eq:dw_final}
    \end{equation}
    which means that the proposed privacy-preserving scheme ensures that the overall difficulty of 
    $N$ sequential aPoW steps is at least as large as the difficulty of a single aPoW step. In addition, 
    a further step to make attacks more difficult is to implement a randomized sub-sampled 
    super-majority-based voting system such that spam attacks become easily visible and mitigating 
    measures can be initiated.
    \newline
    
    \emph{Remark:} Notice that the idea behind the adaptive PoW, while similar in intent to 
    other consensus mechanisms based on {\em reputation}, such as those in \cite{rev1_1,rev1_2}, 
    also differs from these approaches. In particular, rather than simply labelling a certain data entry 
    as untrustworthy, our approach additionally makes it prohibitively costly for an attacker to 
    write malicious information.

\section{Application example - Reinforcement learning over SPToken}\label{sec:estimation_scheme}
    In accordance with the design objectives described in Section~\ref{sec:dlt}, it can be seen that 
    the SPToken architecture is {\em technique-agnostic} in the sense that potentially any 
    population-based technique that solves the dynamic routing problem could be used,
    including not only multi-agent ML but also other methods such as Black-Box Optimization (e.g.,
    evolutionary strategies \cite{stulp2012policy}). However, in order to give a practical 
    illustration of the proposed approach, we will use an RL strategy in combination with the 
    described token-passing architecture. Specifically, instead of using vehicles as RL agents
    \cite{roman_cdc_2019} to probe an unknown environment, we use tokens ``jumping'' among vehicles 
    to effectively create virtual agents and emulate the behaviour of commanded agents designed to 
    probe the surroundings of arbitrary routes. For this, we employ a modified version 
    of the recently proposed RL algorithm called \textbf{U}pper \textbf{B}ounding the 
    \textbf{E}xpected Next State \textbf{V}alue (UBEV)~\cite{ubev}. UBEV involves a combination of 
    {\it backward induction} with maximum likelihood estimation to (i) construct optimistic empirical 
    estimates of state transition probabilities, (ii) assign empirical immediate reward, and (iii) 
    compute optimal policy.
    \newline

    Since long training time is a common disadvantage of RL algorithms, we propose to launch a high 
    number of independent tokens, which act as virtual vehicles and use the same MDP's policy matrix 
    to explore different areas of a city. In fact, our design of the state-action space allows us 
    to avoid estimating the transition probabilities, which significantly reduces the training time. 
    Effectively, the algorithm learns only the reward distribution which describes the environment 
    (e.g., traffic patterns in a city). We also launch multiple virtual agents that collaborate 
    with each other to speed up the learning process. In addition, introducing tokens enables 
    learning by crowdsourcing without biasing the environment: indeed, if one launched a number 
    of agents, each checking a policy, the corresponding part of the environment could have become 
    congested as a result, and such artificial congestion would be learned by the RL algorithm. 
    SPToken allows one to avoid introducing such biases. Thus, in principle, the token-based 
    technique does not require modification of conventional RL algorithms, and can indeed be applied 
    to any machine learning algorithm. Note also that the concept of tokens does not necessarily 
    require using DLT. However, in our approach we use DLT as a supplementary layer to the RL 
    algorithm in order to facilitate robust data management for real-world deployments. Without the 
    DLT layer, critical aspects such as privacy, authentication, security, robustness to spamming, 
    among others, could be compromised. Our particular DLT implementation, SPToken, manages token 
    collections/deliveries as DLT transactions, and data is validated via the ledger's consensus 
    algorithms.
    \newline

    Further details of the proposed approach, together with the corresponding experimental assessment, 
    are provided in the following sections. In particular, we experimentally assess:
    \begin{itemize}
        \item how fast the system learns to avoid traffic jams;
        \item how quickly the system returns to the shortest path policy once the traffic jams clear up;
        \item how the training time varies with respect to the number of independent tokens.\\
    \end{itemize}

    The original UBEV algorithm in~\cite{ubev} performs a standard expectation-maximization trick. 
    Namely, it first fixes the state transition probabilities of the MDP and the expected reward 
    estimates, and uses backward induction to design the optimal deterministic policy which maximizes 
    the expected reward. Next, this policy is used to probe the environment, and the statistics collected 
    over the course of probing are used to update transition probabilities by employing a standard 
    ``frequentist'' maximum likelihood estimator~\cite{HTF}, which simply computes the frequencies of 
    transitioning from one state to another subject to the current action (that can be a function of the 
    current state). Then, the optimal policy (for the updated estimates of the transition probabilities 
    and reward) is recomputed again. This procedure is treated as an \emph{episode} of the training 
    process and is iterated until convergence (as demonstrated in~\cite{ubev}).
    \newline 

    \subsection{Modified UBEV algorithm}
    Recall that an MDP is a discrete-time stochastic control process. Our decision problem is a 
    finite horizon MDP with time horizon $H$, and we assume that the model is known and that the 
    environment is fully observable. An MDP can be represented as a tuple 
    $\langle \S, \A_s, \textbf{P}, \R\rangle$, where
    \begin{itemize}
        \item $\S$ is the set of states, with $|\S| = S$ being the number of states;
        \item $\A_s$ is the set of allowable actions, with $|\A_s| = A_s$ being the number of 
            allowable actions in state $s$, $\A = \bigcup_{s \in \S}\A_s$, and with $|\A| = A$ 
            being the total number of actions;
        \item $\P(s'|s, a, t)$ is the probability of transition from state $s$ under action 
            $a \in \A_s$ to 
            state $s'$ at decision epoch $t\in\H$, $\H=\{1, 2, \dots, H\}$;
        \item $\R(s, a, t)$ is the reward of choosing the action $a \in \A_s$ in the state $s$ at 
            decision epoch $t\in\H$.
    \end{itemize}

    In an MDP, an agent (i.e., the decision maker) chooses action $a_t \in \A_s$ at time $t\in\H$ based 
    on observing state $s_t$, and then receives a reward $r_t$. The trajectory of the MDP is defined 
    as follows: it is assumed that $s_{t+1}\sim\P(\cdot|s_t,a_t, t)$, i.e., the state at time $t+1$ 
    is drawn from a distribution $\textbf{P}$ which depends on $s_t$, $a_t \in \A_s$ and decision 
    epoch $t$. In this case, the total expected reward associated to the policy $\pim:\S\to\A$ is 
    defined as
    \begin{multline}
        \label{eq:MDP_expected_reward}
        \rho(\pim)\coloneqq \E_{s_1\dots s_H}\left[\sum_{t\in\H}\R(s_t,\pim(s_t,t),t)\right] \\
        = \sum_{s\in\S} \P_0(s) V^{\pim}_1(s)\,,
    \end{multline}
    where $\P_0$ is the distribution of the initial state, and $V^{\pim}_t$ is the value function from 
    decision epoch $t$ for policy $\pi$, formally defined as follows:
    
    \begin{multline}
        \label{eq:VF_MDP}
        V_t^{\pim} (s) \\ = \R(s,\pim(s,t),t) + \sum_{s'\in\S} \P(s'| s,\pim(s,t),t) V_{t+1}^{\pim} (s'), 
        \\ \quad V_{H+1}^{\pim}\coloneqq 0.
    \end{multline}
    
    Then, the goal of an agent is to find an optimal trajectory which maximizes the expected reward 
    (\ref{eq:MDP_expected_reward}), and the 
    optimal MDP policy (i.e., the policy maximizing Equation~(\ref{eq:MDP_expected_reward})) is 
    calculated through the backward induction process given by:
    \begin{equation}\label{eq:MDP-optimal-policy}
        \begin{split}
            \pim(s,t)&=\argmax_{a\in\A_s} \left\{\R(s,a,t) \small{+} \sum_{s'\in\S} \P(s'| s,a,t) 
            V_{t+1}^{\pim} (s')\right\}\\
            \pim(s,H)&=\argmax_{a\in\A_s} \R(s,a,H).
        \end{split}
    \end{equation}
    
\begin{figure}[h]
        \centering
        \includegraphics[width=0.5\columnwidth]{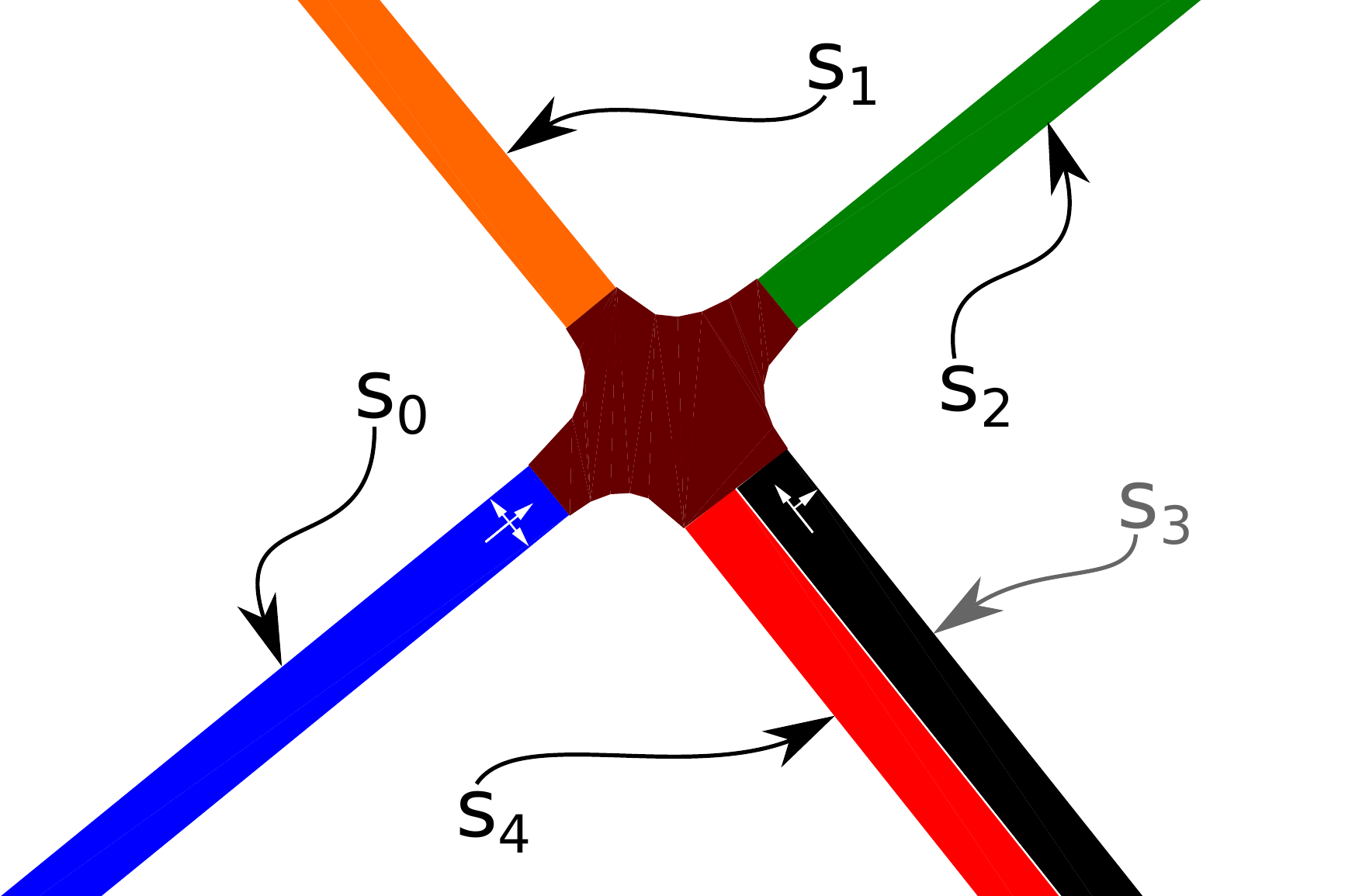}
        \caption{A piece of a road network with five road links representing states 
            $s_0$, $s_1$, $s_2$, $s_3$, $s_4$. Note that 
            state $s_3$ is marked in gray since it is not accessible from 
            state $s_0$.}
        \label{fig_actions}
    \end{figure}

    We are now in a position to present the \textbf{M}odified \textbf{UBEV} (MUBEV) algorithm. 
    Algorithm \ref{alg:modified_ubev} represents a modified version of the UBEV algorithm as a result 
    of adapting the original UBEV algorithm to our target problem, which includes the following 
    modifications to the UBEV algorithm. First of all, we use a specific state-action space.
    A state is represented as a collection of road links, while the action space 
    is based on possible directions of connections between the edges of a road 
    network\footnote{In this work, we do not consider lane-changing behaviour for the agents on 
    multi-lane roads.}. Namely, we apply 
    the actions\ \textquotesingle {\bf s}\textquotesingle---go straight, 
    \textquotesingle{\bf l}\textquotesingle---turn left, 
    \textquotesingle{\bf L}\textquotesingle---turn partially left, 
    \textquotesingle{\bf R}\textquotesingle---turn partially right, 
    \textquotesingle{\bf r}\textquotesingle---turn right, 
    and \textquotesingle {\bf u}\textquotesingle---stay in the same state 
    (which prevents leaving the destination state). We also exclude U-turns to favor the exploration 
    of the environment, as U-turns may result in undesirable recurrent attempts to use the shortest 
    path policy. The proposed model of the state-action space, based on road links and their 
    interconnections, allows us to provide the algorithm with the set of predefined
    trivial transition probabilities. For example, let us construct stochastic rows of transition 
    matrices for all possible transitions from state $s_0$ assuming that actions 
    \textquotesingle {\bf s}\textquotesingle, 
    \textquotesingle {\bf l}\textquotesingle, and \textquotesingle {\bf r}\textquotesingle\ 
    are allowable in state $s_0$ as shown in Figure \ref{fig_actions}. This results in
    $$
    \scriptsize{
    \begin{array}{cc}
        \left[
        \begin{array}{c|>{\centering\arraybackslash$} p{0.1cm} <{$}
                        >{\centering\arraybackslash$} p{0.1cm} <{$}
                        >{\centering\arraybackslash$} p{0.1cm} <{$}
                        >{\centering\arraybackslash$} p{0.1cm} <{$}
                        >{\centering\arraybackslash$} p{0.1cm} <{$}
                        >{\centering\arraybackslash$} p{0.2cm} <{$}}
            \text{{\bf s}} & s_0 & s_1 & s_2 & s_3 & s_4 & \cdots \\
            \hline
            s_0 & 0 & 0 & 1 & 0 & 0 & \cdots \\
            \vdots & \vdots & \vdots & \vdots & \vdots & \vdots & \ddots
        \end{array}
        \right], &
    
        \left[
        \begin{array}{c|>{\centering\arraybackslash$} p{0.1cm} <{$}
                        >{\centering\arraybackslash$} p{0.1cm} <{$}
                        >{\centering\arraybackslash$} p{0.1cm} <{$}
                        >{\centering\arraybackslash$} p{0.1cm} <{$}
                        >{\centering\arraybackslash$} p{0.1cm} <{$}
                        >{\centering\arraybackslash$} p{0.2cm} <{$}}
            \text{{\bf l}} & s_0 & s_1 & s_2 & s_3 & s_4 & \cdots \\
            \hline
            s_0 & 0 & 1 & 0 & 0 & 0 & \cdots \\
            \vdots & \vdots & \vdots & \vdots & \vdots & \vdots & \ddots
        \end{array}
        \right],
    \end{array}
    }
    $$
    $$
    \scriptsize{
    \begin{array}{cc}
        \left[
        \begin{array}{c|>{\centering\arraybackslash$} p{0.1cm} <{$}
                        >{\centering\arraybackslash$} p{0.1cm} <{$}
                        >{\centering\arraybackslash$} p{0.1cm} <{$}
                        >{\centering\arraybackslash$} p{0.1cm} <{$}
                        >{\centering\arraybackslash$} p{0.1cm} <{$}
                        >{\centering\arraybackslash$} p{0.2cm} <{$}}
            \text{{\bf r}} & s_0 & s_1 & s_2 & s_3 & s_4 & \cdots \\
            \hline
            s_0 & 0 & 0 & 0 & 0 & 1 & \cdots \\
            \vdots & \vdots & \vdots & \vdots & \vdots & \vdots & \ddots
        \end{array}
        \right], &
        
        \left[
        \begin{array}{c|>{\centering\arraybackslash$} p{0.1cm} <{$}
                        >{\centering\arraybackslash$} p{0.1cm} <{$}
                        >{\centering\arraybackslash$} p{0.1cm} <{$}
                        >{\centering\arraybackslash$} p{0.1cm} <{$}
                        >{\centering\arraybackslash$} p{0.1cm} <{$}
                        >{\centering\arraybackslash$} p{0.2cm} <{$}}
            \text{{\bf u}} & s_0 & s_1 & s_2 & s_3 & s_4 & \cdots \\
            \hline
            s_0 & 1 & 0 & 0 & 0 & 0 & \cdots \\
            \vdots & \vdots & \vdots & \vdots & \vdots & \vdots & \ddots
        \end{array}
        \right].
    \end{array}
    }
    $$
    Clearly, it is not required to learn such trivial transition probabilities, which is a significant 
    advantage especially for large road networks. Note that, in our model, the action 
    \textquotesingle {\bf u}\textquotesingle\ (stay in the same state) is allowable at each 
    state $s\in\S$.
    \newline

    The second modification addresses the following situation. At the beginning of the training, 
    there is a little or no information at all of the reward distribution, and the algorithm rather 
    explores than exploits. By default, the original algorithm always selects the first component of 
    $Q$ (Algorithm 1, line 25) if all the components of the $Q$-function are equal (Algorithm 1, 
    line 22), and thus it probes the environment without any preference in term of the direction 
    of the exploration. In contrast, we force it to stick to the shortest path policy whenever 
    $Q_i=Q_j$, $\forall i, j$, so that it explores the surrounding along the shortest route.
    Once the agent faces a traffic jam after a certain action $a \in \A_s$, it gets delayed, 
    which in turn introduces the negative reward for the action $a \in \A_s$ at state $s$. As a 
    result, the reward distribution changes, and the shortest path policy is amended to avoid the 
    jam by looking for a detour. By operating in this fashion, we sample along near optimal 
    trajectories which has also a practical value.
    \newline

    \begin{algorithm}[h]
        \caption{Modified \textbf{U}pper \textbf{B}ounding the \textbf{E}xpected Next State 
            \textbf{V}alue (\textbf{UBEV}) Algorithm - MUBEV
        \label{alg:modified_ubev}}
        \begin{algorithmic}[1]
            \Require{$\S$; $\A$; $H$; \textbf{P}; $\Pi_{SP}$; $M$; $\delta \in \left(0, 1\right]$; 
                $r_{max}$.}
            \State $n(s, a, t)=R(s, a, t)=0;\,\hat{V}(s, t')=0;\,\hat{Q}(s, a, t')=0$
            \Statex $\forall s, s' \in \S,\,\, a \in \A_s, t \in \H,\,\, t' \in \{1, 2, \dots, H + 1\}$.
            \State $\delta' = \delta / 9$; $V_{max} = H*r_{max}$.
            \Statex
            \For{$k=1,2,3...$}{}
                \Statex
                \textit{{\color{blue}\ \ \ \ \ // Optimistic planning loop}}
                \For{$t=H$\,\,\textbf{to}\,\,$1$}{}
                    \State $\hat{V}_{t+1} = \hat{V}(\cdot, t+1)$
                    \State $\tilde{V}_{max} = \min\big(\max(\hat{V}_{t+1}), V_{max}\big)$
                    \For{$s \in \S$}{}
                        \For{$a \in \A_s$}{}
                            \State $r = r_{max}$; $EV = \tilde{V}_{max}$
                            \If{$n(s, a, t) > 0$}
                                \State $\eta_{1} = 2\ln\ln\left(\max\left(e, n(s, a, t)\right)\right)$
                                \State $\eta_{2} = \ln\left(18*S*A_s*H/\delta'\right)$
                                \State $\phi = \sqrt{\frac{\eta_{1} + \eta_{2}}{n(s, a, t)}}$
                                \State $\hat{V}_{next} = P(\cdot, s, a, t) \times \hat{V}_{t+1}$
                                \State $EV \small{=} \min\left(\tilde{V}_{max}, \hat{V}_{next} +
                                    \left(H-t\right)\phi\right)$
                                \State $\hat{r}(s, a, t) = \frac{R(s, a, t)}{n(s, a, t)}$
                                \State $r = \min\left(r_{max}, \hat{r}+ \phi\right)$
                            \EndIf
                            \State $Q(a) = r + EV$
                        \EndFor
                        \State $\hat{Q}(s, \cdot, t) = Q$;
                        \If {$Q_i=Q_j\,\,\forall Q_i,Q_j \in Q$}
                            \State $\tilde{a} = \Pi_{SP}(s, t)$
                        \Else
                            \State $\tilde{a} = \argmax_{a\in\A_s}Q(a)$
                        \EndIf
                        \State $\pi_{k}(s, t) = \tilde{a}$; $\hat{V}(s, t) = Q(\tilde{a})$
                    \EndFor
                \EndFor
                \Statex
                \textit{{\color{blue}\ \ \ \ \ // Execute policy for one episode}}
                \State $\tilde{s} = \big[s_1^{(1)}, ..., s_1^{(M)}] \small{\sim}
                    \mathcal{U}\big(1, S\big),\ s_1^{(i)} \small{\neq} s_1^{(j)}\ \forall i, j 
                    \small{\in}[1, M]$
                \For{$t=1$\,\,\textbf{to}\,\,$H$}{}
                    \State $a_t^{(m)} = \pi_k(s_t^{(m)}, t)$
                    \State $s_{t+1}^{(m)} \sim  P(\cdot \vert s_t^{(m)}, a_t^{(m)}, t)$
                    \State $r_t = \R\big(s_t^{(m)}, s_{t+1}^{(m)})$
                        \Comment{{\color{blue} Call to Function 1}}
                    \State $R\big(s_t^{(m)}, a_t^{(m)}, t'\big) += r_t,\,\,  \forall t' \in \H$
                    \State $n\big(s_t^{(m)}, a_t^{(m)}, t'\big)++,\,\, \forall t' \in \H$
                \EndFor
            \EndFor
        \end{algorithmic}
    \end{algorithm}
    
    Concerning the third modification, we aim to launch multiple tokens always starting at different 
    (randomly sampled) origins and having the same destination. All these tokens follow the same 
    MDP's policy matrix, and the corresponding collected statistics are then used to update the policy. 
    Therefore, learning and adaptation happen more rapidly.
    \newline
    
    Finally, we propose a stationary model of the MDP (i.e., transition probabilities and the reward 
    distribution do not vary with time) where each independent agent (token) contributes to the MDP's 
    reward matrix, and they all use new updated policy in the next episode of the learning process.
    \newline

    \begin{algorithm}[h]
        \caption*{\textbf{Function 1} The Reward Function
        \label{alg:reward}}
            \begin{algorithmic}[1]
                \Require{$s_t$; $s_{t+1}$; $\tau(s_{t+1})$; $\alpha$; $\beta$; $w_D$; $w_T$; $\Omega$;
                $r_{max}$.}
                \Ensure{$r_t$.}
                \Function{$\R$}{$s_t$; $s_{t+1}$}
                    \If {$s_{t+1} \neq s_t$}
                        \Statex
                        \textit{{\color{blue}\ \ \ \ \ \ \ \ \ // Distance reward computation}}
                        \State $d = D(s_{t}) - L(s_{t})$
                        \If {$d \neq 0$}
                            \State $r_D = r_{max} - \frac{D(s_{t+1})}{d}$
                        \Else
                            \State $r_D = r_{max}$
                        \EndIf
                        \Statex
                        \textit{{\color{blue}\ \ \ \ \ \ \ \ \ // Time reward computation}}
                        \State $\tau_{ref} = RY(s_{t+1}) + \alpha * \tau_{min}(s_{t+1})$
                        \If {$\tau(s_{t+1}) \leq \tau_{ref}\ \textbf{or}\ s_{t+1} = s_f$}
                            \State $r_T = 0$
                        \Else
                            \State $r_T = - \beta * \frac{\tau(s_{t+1})}{\tau_{ref}}$
                            \Statex
                            \textit{{\color{blue}\ \ \ \ \ \ \ \ \ \ \ \ \ // 
                                Applying the edge coefficient}}
                            \If {$RY(s_{t+1}) = 0$}
                                \State $r_T = r_T * EC(s_{t+1})$
                            \EndIf
                        \EndIf
                        \State $r_t = w_D*r_D + w_T*r_T$ \Comment{{\color{blue}Total reward}}
                    \Else \Comment{{\color{blue}Jumping to the same state}}
                        \If {$s_{t+1} \neq s_f$}
                            \Statex
                            {\it 
                            {\color{blue}\ \ \ \ \ \ \ \ \ \ \ \ \ // 
                                Penalty: staying at not destination}}
                            \State {$r_t = - \Omega$}
                        \Else
                            \State $r_t = r_{max}$
                        \EndIf
                    \EndIf
                    \Return $r_t$
                \EndFunction
        \end{algorithmic}
    \end{algorithm}

    \textbf{Notation for MUBEV and the Reward Function.}
    In Algorithm 1 we have:
    $\S$ is the set of states;
    $\A_s$ is the set of allowable actions in state $s$, $\A = \bigcup_{s \in \S}\A_s$; 
    $S$ and $A$ denote cardinality of finite sets $\S$ and $\A$ respectively;
    $A_s$ denote cardinality of a finite set $\A_s$;
    $H$ is the length of the MDP's time horizon, with $\H = \{1, 2, \cdots, H\}$; 
    \textbf{P} is an array of predefined transition probabilities; 
    $\Pi_{SP}$ is the shortest path policy; 
    $M$ is the number of MUBEV tokens;
    $\delta$ is the failure probability (see \cite{ubev} for details); 
    $n(s, a, t)$ is the number of actions $a \in A_s$ taken from state $s$ at time $t$;
    $R(s, a, t)$ is accumulated reward from state $s$ under action $a \in A_s$ at time $t$;
    $\hat{V}(s, t')$ is the value function from time step $t'$ for state $s$;
    $\hat{Q}(s, a, t)$ is the Q-function for the appropriate state, action and time \cite{ubev}.
    Initial values of elements in arrays $n$, $R$, $\hat{V}$ and $\hat{Q}$
    are zeros for all $s \in \S, a \in \A_s, t \in \H, t' \in \{1, 2, \cdots, H+1\}$. 
    Additionally: $r_{max}$ is the maximum reward that the agent can receive per one transition;
    $V_{max}$ is the maximum value for next states; 
    $\hat{V}(\cdot, t+1)$ and $P(\cdot, s, a, t)$ denote vectors of length $S$, and 
    $\hat{Q}(s, \cdot, t)$ is interpreted as a vector of length $A_s$; 
    $\phi$ is the width of the confidence bound \cite{ubev}; $e$ is the Euler's number;
    $\hat{r}(s, a, t)$ is normalized reward from state $s$ under action $a \in A_s$ at time $t$; and 
    $r$ and $EV$ are auxiliary variables. Vector $\tilde{s}$ is a vector of initial states of MUBEV 
    tokens, which is uniformly sampled in range from $1$ to $S$ with no repeated entries.
    The agents (tokens) interact with the environment each time step $t \in \H$, and receive reward 
    $r_t$ determined by the reward function defined in Function 1.
    \newline

    Concerning Function 1, it returns total reward, i.e., distance reward plus time reward, at time $t$.
    Additionally: $\tau(s_{t+1})$ is actual travel time on edges that correspond to state $s_{t+1}$;
    $\alpha$ is a scale factor that increases minimum travel time on edges due to traffic uncertainties;
    $\beta$ is a parameter used for faster learning of congestions;
    $\omega_D$ and $\omega_T$ are the weights of distance and time reward, respectively;
    $\Omega$ is the absolute value of penalty given to an agent if it takes action \textquotesingle 
    {\bf u}\textquotesingle\ at a state other than the destination state, or when it leaves the 
    destination;
    $D(s_t)$ is the shortest route length from state $s_t$ to the destination state $s_f$; and 
    $L(s_t)$ is the length of edges that correspond to state $s_t$. Finally,
    $RY(s_{t+1})$ is the duration of yellow plus red phases of traffic light signals (TLS) that control 
    edges that represent state $s_{t+1}$;
    if all edges in some state are not controlled by a TLS, we apply $RY = 0$ for that state.
    If some edges are not controlled by traffic light signals, we employ the edge coefficient $EC$
    for them (Function 1, lines 13-14) which is computed in this fashion: 
    if the length $L_{t+1}$ of edges that correspond to state $s_{t+1}$ is smaller than the average 
    length $\bar L$ of edges included in states, then $EC(s_{t+1}) = \left[L(s_{t+1})/\bar L\right]^4$, 
    otherwise  $EC\small = 1$.

\section{Numerical simulations}
    In the following application, we are interested in designing a recommender system for a community of 
    road users. We distribute a set of MUBEV tokens so that the uncertain environment can be ascertained.
    These tokens are passed from vehicle to vehicle using the DLT architecture described in Section 3.
    Specifically, tokens are passed from one vehicle to another in a manner that emulates a real 
    vehicle probing an unknown environment under the instruction of the RL algorithm. The token 
    passing is determined by both the operation of the MUBEV algorithm and the DLT infrastructure, 
    both of which can be orchestrated using a cloud-based service. Vehicles possessing a token are 
    permitted to write data to the DLT. We refer to such vehicles as virtual MUBEV vehicles. In this 
    way, the token passing emulates the behaviour of a real agent (vehicle) that is probing the 
    environment. Once the environment has been learnt, a route recommender system can be built 
    for a community of users interested in route recommendations (e.g., through a smart device app).
    \newline

\begin{figure}[h]
        \centering
        \includegraphics[width=1.0\columnwidth]{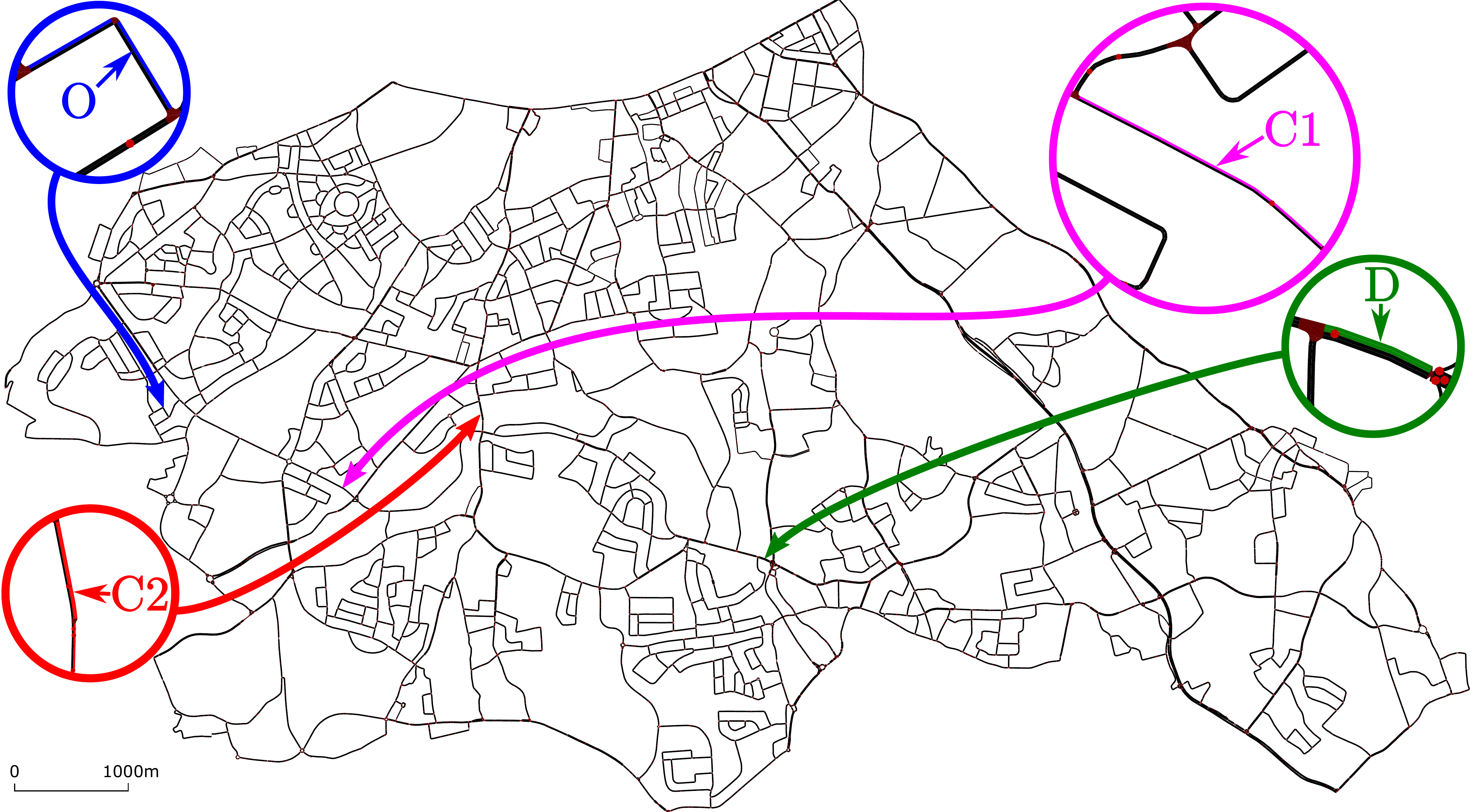}
        \caption{Realistic road network used in the experiments: a part of Dublin, Republic of Ireland. 
            Four road segments of interest are highlighted, namely $O$, $D$, $C1$ and $C2$.}
        \label{fig_exp_setup}
    \end{figure}

\begin{figure}[h]
        \centering
        \includegraphics[width=0.65\columnwidth]{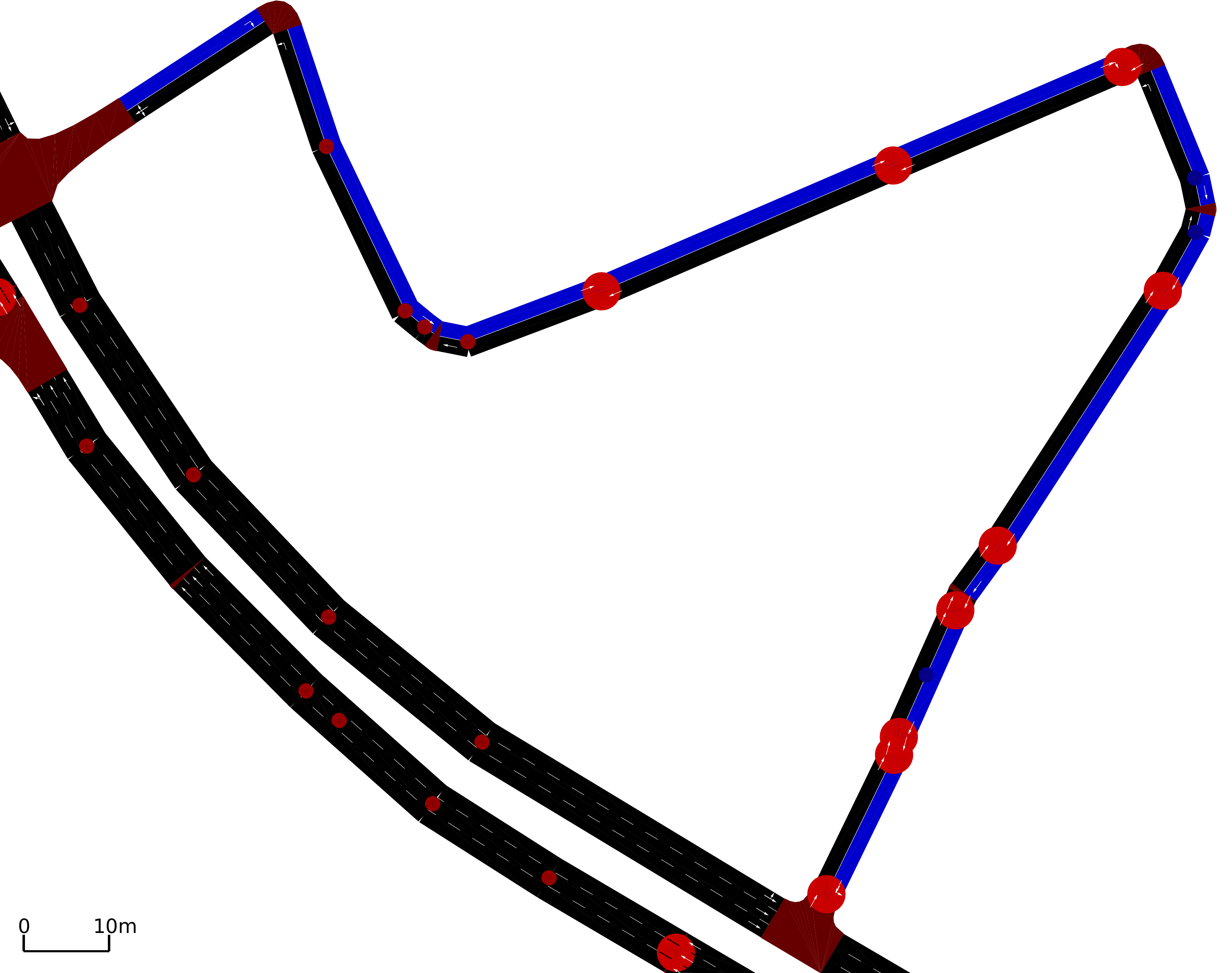}
        \caption{State model: a state corresponds to a set of road links. Road links marked in blue are 
            merged into one state.}
        \label{fig_merging}
    \end{figure}

    For the experimental evaluation of our proposed approach, we designed a number of numerical 
    experiments based on traffic scenarios implemented with the open source traffic simulator SUMO 
    \cite{sumo}. Interaction with running simulations is achieved using Python scripts and the SUMO 
    packages {\it TraCI} and {\it Sumolib}. The general setup used in our simulations is as follows:
    \begin{itemize}
        \item In all our experiments, we make use of the area in Dublin, Republic of Ireland shown in 
            Figure \ref{fig_exp_setup} (roughly speaking, south Dublin city centre to the canal), with 
            all the U-turns removed from the network file, and a total of $S = 3,663$ states.
        \item A number of roads are selected as origins, destinations, and sources of congestion
            (see Figure \ref{fig_exp_setup}). In all experiments we use the set \{$O$, $D$\} as an 
            origin-destination (OD) pair. We use $C1$ in Experiment 1 and 2 and \{$C1$, $C2$\} in 
            Experiment 3 to simulate traffic jams on them.
        \item In all simulations we use a new vehicle type based on the default SUMO vehicle type
            with maximum speed = 118.8 km/h and impatience\footnote{Readiness of a 
            driver to impede vehicles with higher priority. For other parameters, see the full 
            documentation at 
            \url{https://sumo.dlr.de/wiki/Definition_of_Vehicles_Vehicle_Types_and_Routes}.} = 0.5. 
        \item For the generation of traffic jams, we modify the maximum speed of certain cars to be 
            6.12 km/h and populate the selected roads with them. When these vehicles are in possession 
            of a token, they become virtual MUBEV vehicles.
        \item Whenever required, shortest path is obtained via SUMO using the default routing 
            algorithm ({\it dijkstra}).
        \item We refer to the state of an agent as a set of road sections (see Figure \ref{fig_merging}) 
            and to a token trip as a RL episode.
        \item We set $H = 85$ for the length of the MDP's time horizon\footnote{This number corresponds 
            to the length of the largest shortest path between an arbitrary origin and destination $D$ 
            (55 transitions in our case) plus a degree of freedom of 30 transitions to properly cope 
            with uncertainties.}.
    \end{itemize}

    {\bf Comment: }Note that in our previous works \cite{roman_cdc_2019, roman_iccve_2019}, models were 
    developed in which a state corresponded to a single road segment. Such a model is not efficient for 
    number of reasons. First, it leads to prohibitively large RL graphs, even for medium-sized road 
    networks. Second, it also leads to sparse graphs (which can easily be reduced).
    \newline

    To overcome the issues addressed in the above comment, we pre-process the road network to allow 
    the merging of different road links into one state: for a given road link, whenever there is only 
    one incoming direction and only one outgoing direction with neighbor road links, we joint those 
    edges into a single state (as shown in Figure \ref{fig_merging}). With this preparatory step, 
    the map shown in Figure~\ref{fig_exp_setup} that contains 10,803 road links results in a graph 
    with only 3,663 states.

    Concerning the design parameters of the reward funcion and the MUBEV algorithm, in all our 
    experiments we set $\omega_D = \omega_T = 1$, $r_{max} = 1$, $\delta = 1$, and tuned the other 
    design parameters as follows: $\alpha = 1.2$, $\beta = 1.3$, $\Omega = 20$. 
    All the experiments rely on a number of background vehicles with random routes which will 
    potentially carry tokens if required. Any specific additional setup for each individual
    experiment will be described in the corresponding subsection below.

    \subsection{Experiment 1: Optimal route estimation under uncertainty}
    The purpose of the first experiment is to determine if our DLT-enabled RL approach can estimate a 
    simple unknown environment. To this end, we define the following experiment. We specify an OD pair 
    for which shortest path is known and then, at various time instances, artificially introduce 
    congestion along the shortest path. For this scenario, we aim to show that the token-enabled MUBEV 
    algorithm can distinguish between these two situations, and, in the case of congestion, find the 
    next best route between origin and destination.
    \newline

\begin{figure}[h]
        \centering
        \includegraphics[width=1.0\columnwidth]{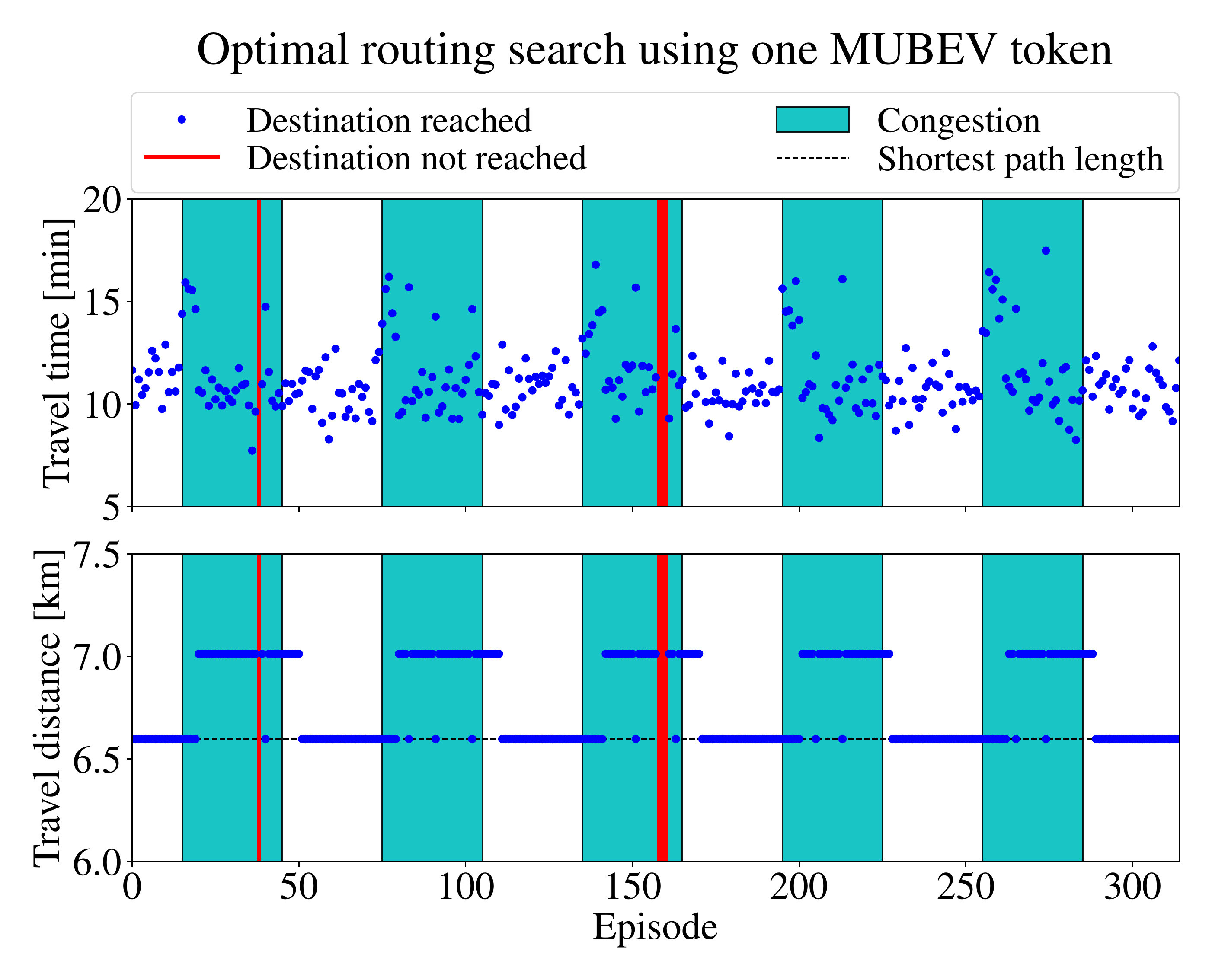}
        \caption{Experiment 1: Travel time and travel distance of a single MUBEV token 
            during the iterative learning process on a changing environment, using a fixed OD pair and 
            approaching an intermittently congested road link ($C1$). Each datapoint corresponds to 
            information registered at the end of each episode (i.e. trip).}
        \label{fig_exp1}
    \end{figure}

    Specifically, this first experiment is conducted as follows. We use a single token over each episode 
    of the learning process, meaning that data from the token is used to update the MUBEV policy over 
    every single episode. For this, the MUBEV token has a fixed OD pair given by \{$O$, $D$\}, and we 
    select the road section labeled as $C1$ in Figure \ref{fig_exp_setup} (which belongs to the 
    shortest path for the selected OD pair) to generate a traffic jam on it at different intervals. 
    Then, over each new episode we start the token from $O$ and ask it to travel to $D$, keeping a 
    record of its performance in terms of travel distance (route length) and travel time regardless 
    of its success in attempting to reach $D$. Additionally, the token has a maximum number of allowed 
    links (defined by the length of the MDP's time horizon) that it can traverse, and if it does not 
    reach its destination within this restriction, then the token trip is declared incomplete 
    (i.e. unsuccessful). The results for this experiment are shown in Figure \ref{fig_exp1}, from 
    which we can draw two main conclusions:
    \begin{itemize}
        \item in general, we can see that the token succeeds in both avoiding traffic jam once 
            congestion is created, and returning to shortest path once congestion is removed, using
            a reasonably small number of episodes (see Figure \ref{fig_exp1} bottom);
        \item as time passes, more statistical information is collected from the environment in the form 
            of reward, and the token is more likely to fully complete a trip for the given OD pair 
            (i.e. red stripes eventually disappear as the experiment progresses in Figure 
            \ref{fig_exp1}).
            \newline
    \end{itemize}

    These two observations validate our expectations about the UBEV-based routing system: (i) it is 
    able to adapt to uncertain environments, and (ii) its performance improves as time passes. It is 
    worth noting that this experiment is useful to analyze the performance of a single token in the 
    iterative learning process from the environment using a fixed OD pair. 

    \subsection{Experiment 2: Optimal route planning under multiple uncertainties}
\begin{figure}[h]
        \centering
        \includegraphics[width=1.0\columnwidth]{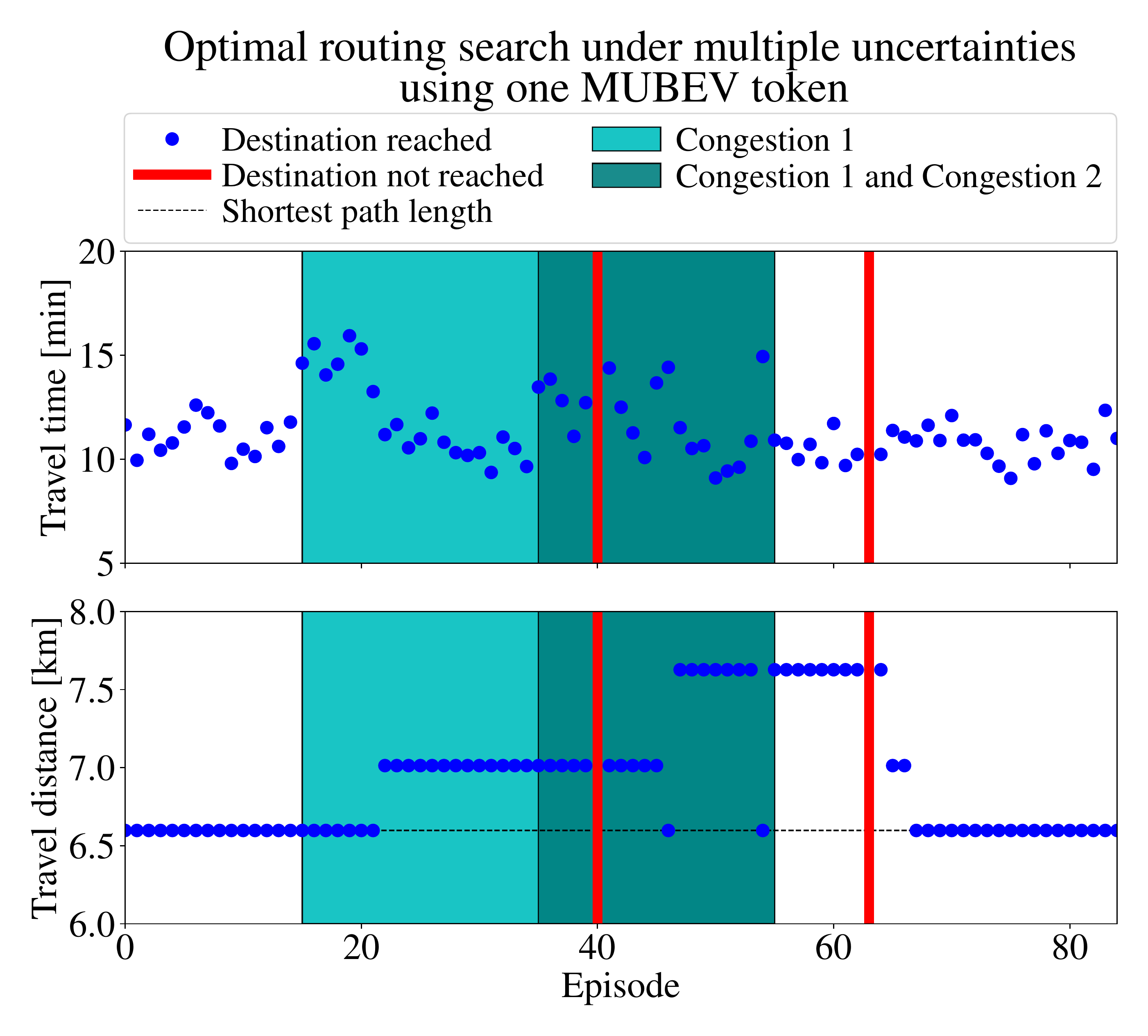}
        \caption{Experiment 2: Travel time and travel distance of a single MUBEV token 
            using a fixed OD pair during the iterative learning process with multiple 
            uncertainties on the environment, namely two intermittently congestions are introduced 
            on $C1$ and $C2$. Each datapoint 
            corresponds to information registered at the end of each episode (i.e. trip).}
        \label{fig_two_jams}
    \end{figure}

    Now we want to evaluate optimal routing under multiple uncertainties. For this,
    we use a similar setup to that in Experiment 1 (i.e., same OD pair, intermittent traffic jam on 
    $C1$, and one MUBEV token probing the environment), and additionally include a second traffic 
    jam using the following procedure: 1) traffic jam on $C1$ is introduced, 2) the system learns the 
    optimal detour, and 3) second traffic jam is introduced on such an optimal detour (specifically on
    road link $C2$, as seen in Figure~\ref{fig_exp_setup}). The results of a single realization of this 
    procedure are shown in Figure~\ref{fig_two_jams}.

    As it can be seen in Figure~\ref{fig_two_jams}, a new optimal detour can be learnt after the 
    second traffic jam is created, and the system rapidly returns to the shortest path policy once 
    the two congestions are removed. Recall that red stripes represent uncompleted routes (i.e., the 
    destination state is not reached within an episode), which are more likely to appear once two 
    different congestions are faced.
    \newline

    Note that once the environment has been determined, the resulting route recommendations gleaned 
    from the UBEV-based system can be made available to the wider community of vehicles. We explore 
    this in the following experiment.
    \newline

    \subsection{Experiment 3: Route recommendations from the UBEV-based system and speedup in learning}
\begin{figure}[h]
        \centering
        \includegraphics[width=1.0\columnwidth]{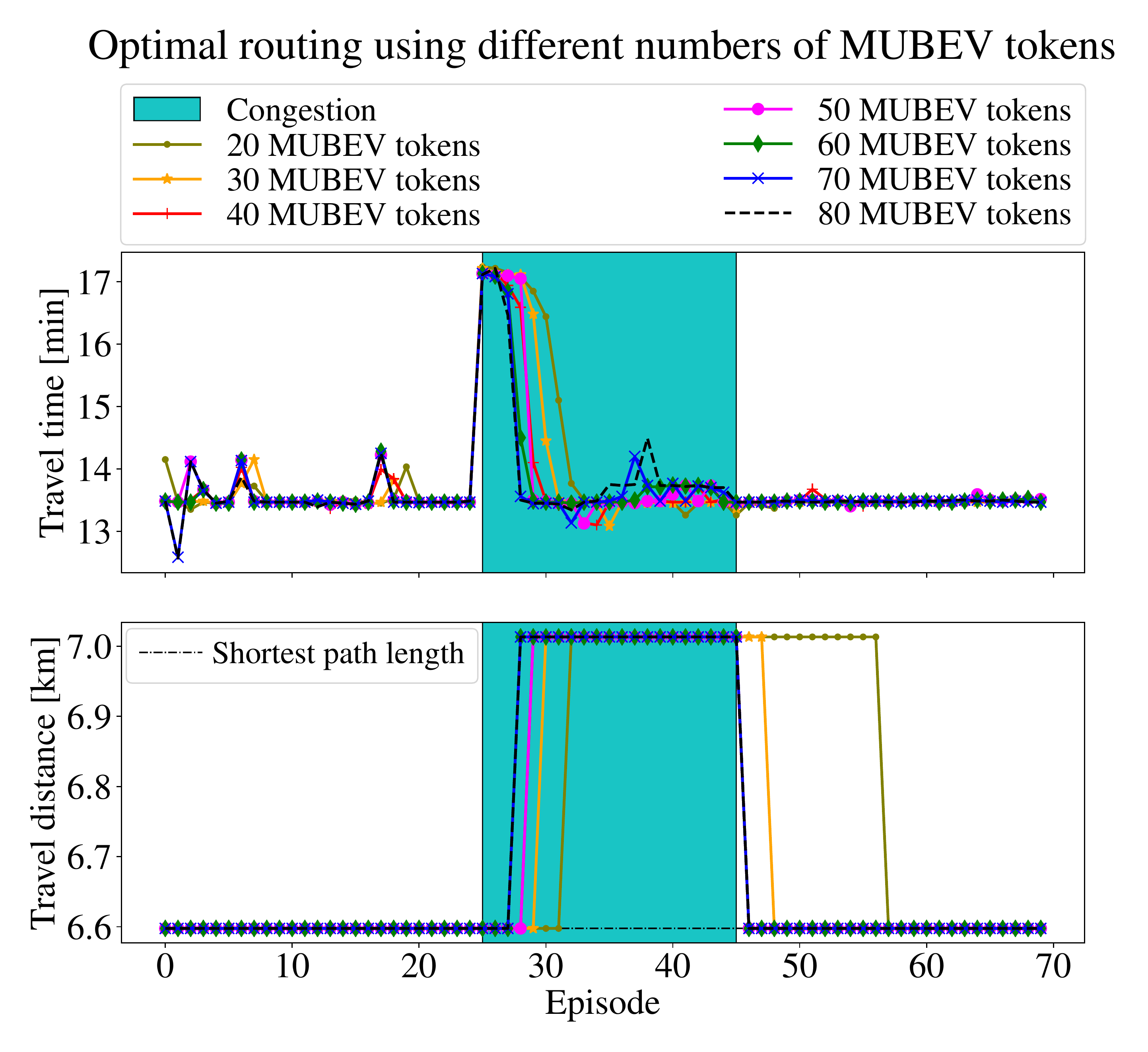}
        \caption{Experiment 3: Median values of travel time/distance of a test vehicle using route 
            recommendations from a UBEV-based routing system involving multiple MUBEV tokens. 
            Each datapoint corresponds to the median value of 250 different realizations of the 
            experiment, with travel time/distance collected at the end of each episode (i.e. trip).}
        \label{fig_exp3_median}
    \end{figure}

\begin{figure}[h]
        \centering
        \includegraphics[width=1.0\columnwidth]{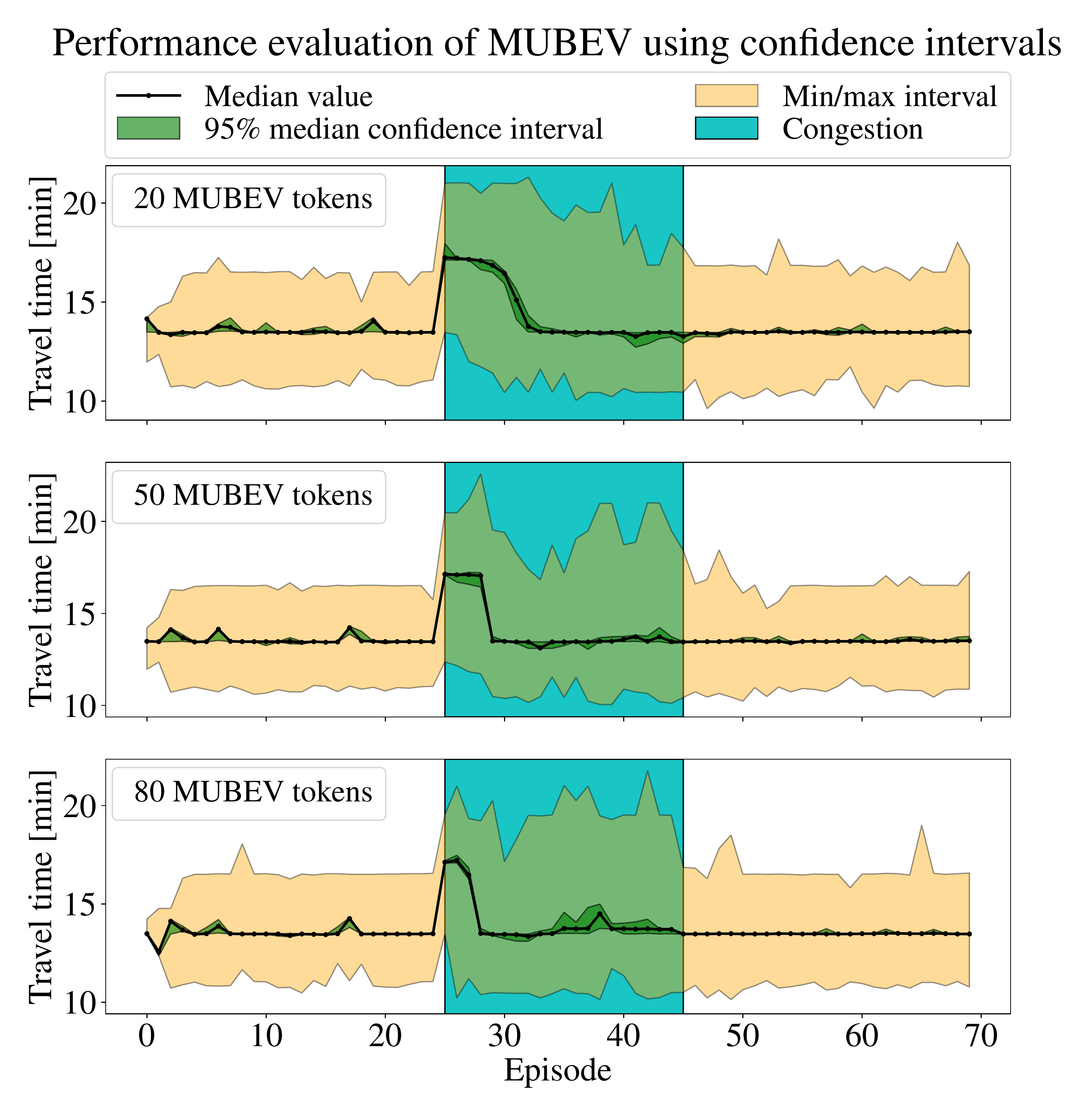}
        \caption{Experiment 3: 95\% confidence intervals for the median travel time when using 20, 50 
            and 80 MUBEV tokens. Each datapoint reflects the analysis per episode of the 250 different 
            realizations of the experiment, with travel time collected at the end of each episode.}
        \label{fig_exp3_confint}
    \end{figure}

\begin{figure}[H]
        \centering
        \includegraphics[width=1.0\columnwidth]{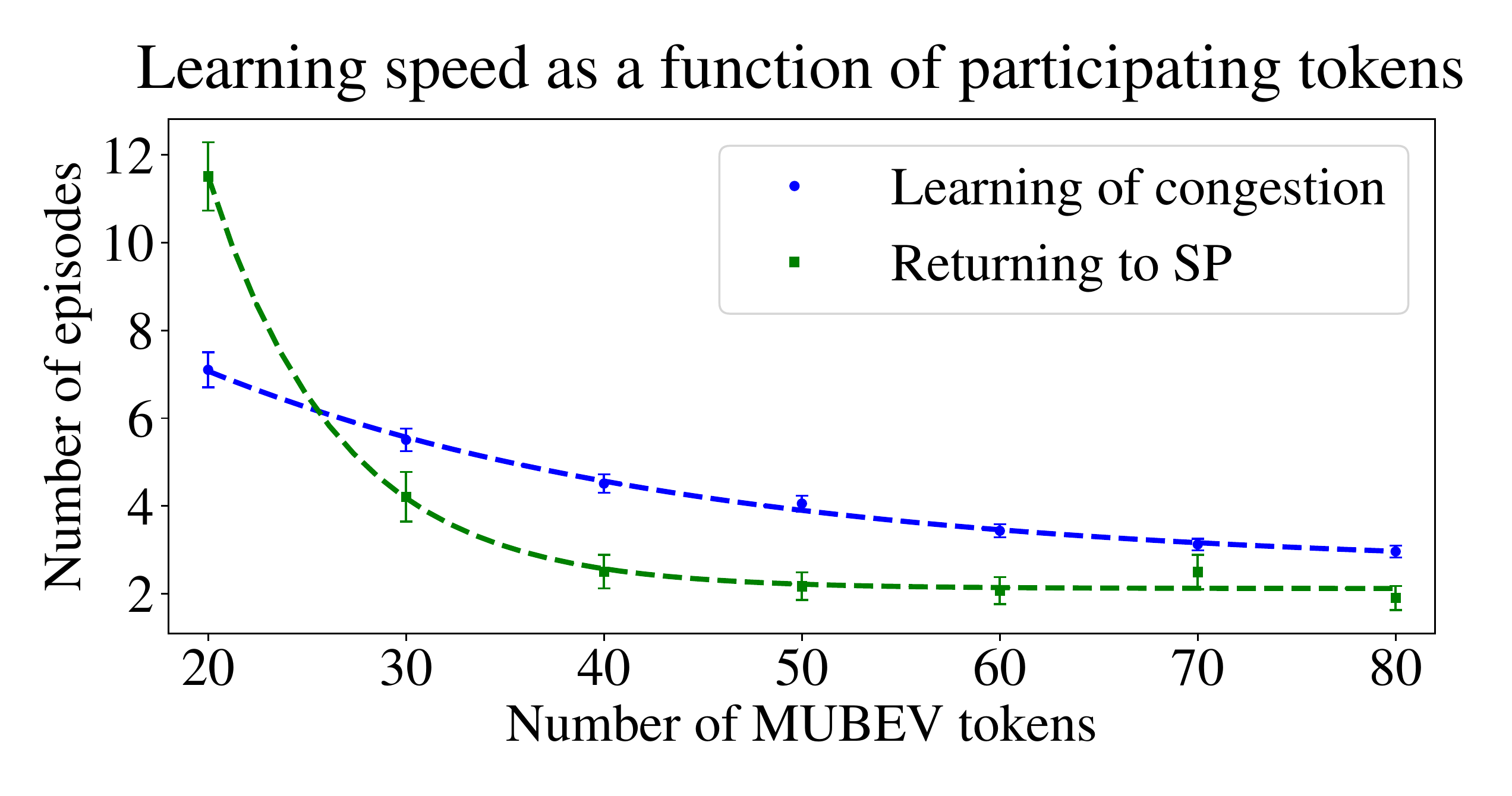}
        \caption{Experiment 3: Average learning speed using multiple MUBEV tokens.
            Each datapoint corresponds to the average value obtained for 250 different realizations 
            of the experiment. [The error bars represent the 95\% confidence 
            interval for the mean. Dashed lines were obtained using exponential curve fitting.]}
        \label{fig_exp3_speedup}
    \end{figure}

    The previous experiments are a simple demonstration of the use of MUBEV in a mobility 
    context. We now explore a scenario where multiple tokens, starting from different origins, are used 
    to update the MDP's policy over each episode. Specifically, in this third experiment, we evaluate the 
    performance of MUBEV as a function of the number of tokens over each episode, subject to a 
    uniform spatial distribution of origins and a common destination (namely, road link $D$). 
    Additionally, we analyze the performance of a {\it test} (non-MUBEV) {\it vehicle} 
    trying to reach destination $D$ from the given fixed origin $O$, using a recommendation from a 
    simplistic UBEV-based routing system. In this case, the initial recommendation corresponds to the 
    shortest path policy, and further recommendations come from the refinement of such a policy. 
    In addition, if a complete route cannot be calculated using the MUBEV recommender system, then the 
    most recent valid recommendation is reused. Remember that the MDP's policy is updated at the end 
    of each episode, and thus we only release a new test vehicle at the end of each episode (once the 
    policy has been updated). The results for this experiment, obtained from a total of 250 realizations 
    in order to have a high statistical significance, are depicted in Figures 
    \ref{fig_exp3_median}, \ref{fig_exp3_confint} and \ref{fig_exp3_speedup}.

    In Figure~\ref{fig_exp3_median}, it can be observed that the number of participating tokens directly 
    affects the convergence rate of the algorithm. As expected, the more tokens involved, the faster the 
    learning process. Figure \ref{fig_exp3_confint} shows the corresponding 95\% median confidence
    interval (MCI)\footnote{The interpretation of the MCI is analogous to the Confidence Interval in 
    the sample mean case. However, the results derived from the MCI analysis have higher accuracy 
    \cite{strelen2004accuracy}.} when using 20, 50 and 80 tokens, respectively, which reflects the high 
    reliability (i.e. narrow MCI range) of our proposed method. From Figure~\ref{fig_exp3_speedup}, 
    we can notice more conclusively the relationship between the number of MUBEV tokens and the average 
    number of episodes required to learn a new given traffic condition (either congestion or free 
    traffic), which reflects an exponential-like decay.

    \subsection{Experiment 4: SPToken architecture using other optimal policy search algorithm 
    different than MUBEV}
\begin{figure}[h]
        \centering
        \includegraphics[width=1.0\columnwidth]{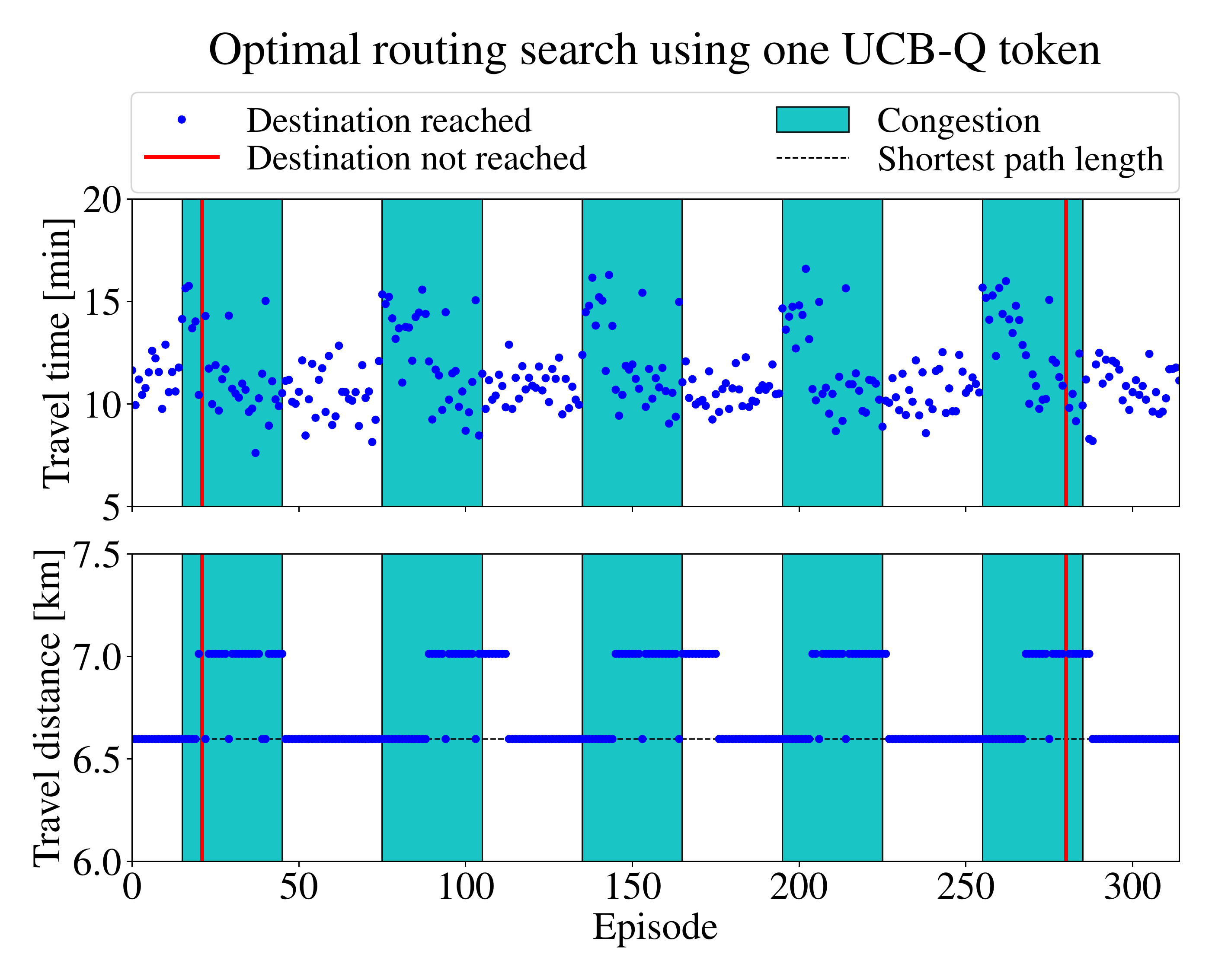}
        \caption{Experiment 4: Travel time and travel distance of a single Q-Learning token, 
            using the setup as in Experiment 1.}
        \label{fig_exp4a}
    \end{figure}

\begin{figure}[h]
        \centering
        \includegraphics[width=1.0\columnwidth]{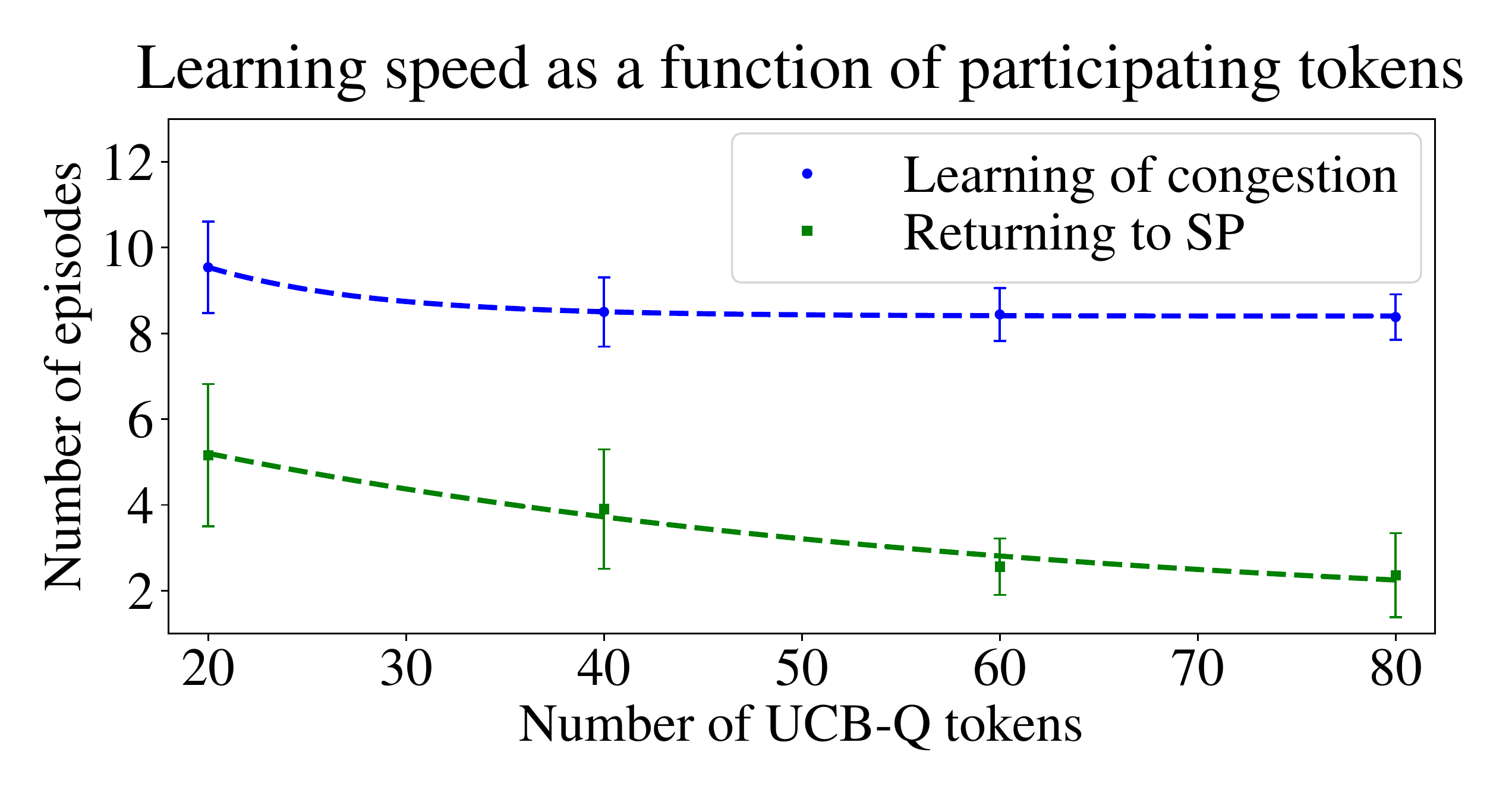}
        \caption{Experiment 4:  Average learning speed using multiple Q-Learning tokens.
            Each datapoint corresponds to the average value obtained at the end 
            of 50 different realizations of the experiment. [The error bars represent the 
            95\% confidence interval for the mean. Dashed lines were obtained using exponential
            curve fitting.]}
        \label{fig_exp4b}
    \end{figure}

    The exhaustive list of algorithms for optimal policy search supported by the SPToken architecture 
    does not only include model-based episodic methods (such as UBEV and related) \cite{ubev, 
    stulp2012policy}, but also model-free episodic methods (such as episodic Q-Learning) 
    \cite{stulp2012policy, blundell2016model, jin2018q}, and Black-Box Optimization methods (such as 
    Natural Evolution and Neuroevolution strategies) \cite{stulp2012policy, pritzel2017neural}. Since 
    evaluating all the possible alternatives is beyond the scope of this paper, we simply illustrate 
    the generality of our approach by exploring the performance of the system when Q-Learning with 
    Upper-Confidence Bound exploration strategy (UCB Q-Learning) \cite{jin2018q} is used instead of 
    MUBEV. The outcome of this is depicted in Figures \ref{fig_exp4a} and \ref{fig_exp4b}, obtained 
    from a total of 50 realizations in order to have a reasonable statistical significance for 
    illustration purposes. The results obtained for UCB Q-Learning in the reduced-scale scenario of 
    Experiment 1 (Figure \ref{fig_exp4a}) show similarities with respect to Figure \ref{fig_exp1} 
    (where MUBEV is used): it is clear that the system is able to learn the optimal policy under both 
    free and congested traffic, although UCB Q-Learning falls behind MUBEV in the latter case. The 
    amplified effects of this feature are more evident when analyzing the performance of UCB Q-Learning 
    in the large-scale scenario described in Experiment 3, with Figure \ref{fig_exp4b} displaying 
    exponential-like speedups but with overall improvements less remarkable than the ones obtained with 
    MUBEV. It is worth noting that these performance issues could not only result from the nature of the 
    algorithm employed for the optimal policy search (e.g., whether it is model-free or model-based), 
    but also depend on factors like a comprehensive calibration process, the number of design parameters 
    to be tuned, and the design of a suitable reward function, among others.

    \section{Conclusion and Outlook}\label{sec:conclusion}
    We introduced a distributed ledger technology design for smart mobility applications.
    As a use case of the proposed approach, we presented a DLT-supported distributed RL algorithm to 
    determine an unknown distribution of traffic patterns in a city. Future work will evolve in two 
    directions. We are currently building an SPToken network using our test vehicles at University 
    College Dublin. Our immediate objective is using this framework to test and prototype applications 
    that consume crowdsourced mobility data. A second avenue of research will be theoretical and 
    related to the network delays associated with transactions over a distributed ledger. The impact 
    of such delays on MUBEV and other optimal policy search algorithms are yet to be quantified.

    \section*{Acknowledgements}
    This work was partially supported by SFI grant 16/IA/4610, 
    and it has been carried out using the ResearchIT Sonic cluster which was funded by UCD IT 
    Services and the Research Office.
    This work is also part funded by IOTA foundation.

    \bibliography{refs}
    \bibliographystyle{ieeetr}

\end{document}